\documentclass{ws-ijbc}
\usepackage{ws-rotating}     
\begin{document}

\catchline{}{}{}{}{} 

\markboth{R. Suresh et al.}{Global and partial phase synchronizations in arrays of time-delay systems}

\title{GLOBAL AND PARTIAL PHASE SYNCHRONIZATIONS IN ARRAYS OF PIECEWISE LINEAR TIME-DELAY SYSTEMS}

\author{R.~Suresh$^{1}$, D.~V.~Senthilkumar$^{2}$, M.~ Lakshmanan$^{1,}$\footnote{Author for correspondence}, and J.~Kurths$^{2,3,4}$}

\address{$^1$Centre for Nonlinear Dynamics, School of Physics, Bharathidasan University,\\ Tiruchirapalli - 620 024, India\\ 
$^{*}$lakshman@cnld.bdu.ac.in}
\address{$^2$Potsdam  Institute for Climate Impact Research, 14473 Potsdam, Germany}
\address{$^3$Institute of Physics, Humboldt University, 12489 Berlin, Germany} 
\address{$^4$Institute for Complex Systems and Mathematical Biology, University of Aberdeen, Aberdeen AB24 3UE, United Kingdom}

\maketitle

\begin{history}
\received{(to be inserted by publisher)}
\end{history}


\begin{abstract} 
In this paper, we report the phenomena of global and partial phase synchronizations in linear arrays of 
unidirectionally coupled piecewise linear time-delay systems.
In particular, in a linear array with open end boundary conditions, global phase synchronization (GPS)
is achieved by a sequential synchronization of local oscillators in the array 
as a function of the coupling strength (a second order transition). 
Several phase synchronized clusters are also formed during the
transition to GPS at intermediate values of the coupling strength, as a prelude to
full scale synchronization. 
On the other hand, in a linear array with closed end boundary
conditions (ring topology), partial phase synchronization (PPS) is achieved 
by forming different groups of phase synchronized clusters above some threshold value of the
coupling strength (a first order transition) where they continue to be in a stable PPS state. 
We confirm the occurrence of both global and partial 
phase synchronizations in two different piecewise
linear time-delay systems using various qualitative and quantitative measures
in three different framework, namely, using explicit phase, recurrence quantification
analysis and the framework of localized sets.  
\end{abstract}


\section{\label{sec:level1}Introduction} 

Chaotic phase synchronization (CPS) associated with a locking of the phases 
of coupled chaotic systems corresponds to the case where the amplitudes are still uncorrelated. CPS
has been investigated in ensembles of globally 
coupled arrays \cite{pikovsky01, boccaletti02, ivanchenko04, takamatsu00, pikovsky96, 
kiss02, zhou02, osipov97, zhan00, kozyreff00, otsuka06}, networks of oscillators 
\cite{boccaletti06,arenas08, batista07, ren07, yu09}, laser systems \cite{kozyreff00, otsuka06}, 
cardiorespiratory systems \cite{schafer09, stefanovska00, bartsch07}, 
ecology \cite{blasius99, sismondo90, amritkar06},
climatology \cite{rybski06, yamasaki09, maraun05}, etc. 
CPS is well studied and understood in low dimensional systems; however, there exist a very little 
indepth studies in higher dimensional systems such as time-delay systems, 
which are essentially infinite-dimensional and exhibit highly non-phase-coherent 
hyperchaotic attractors with complex topological structures. So estimating the phase explicitly
to identify phase synchronization in such systems is quite difficult. Recently,
the occurrence of phase synchronization in time-delay systems has been reported
\cite{senthilkumar06, senthilkumar08}. However, these
investigations are carried out so far only in a system of two coupled time-delay systems. Very recently, 
the occurrence of global phase synchronization in a linear array of Mackey-Glass time-delay systems
in the chaotic regime has been reported in Suresh {\it {et al.,}} [2010] 
 for open end boundary conditions. 
It has been shown that global phase synchronization occurs via a sequential type synchronization 
as the coupling strength increases. To verify the generic nature of the results of 
Suresh {\it {et al.,}} [2010], 
we consider in this paper two different piecewise linear time-delay systems with complex topological structures 
and investigate the occurrence of global phase synchronization along with the underlying mechanism 
for open end boundary conditions. In addition we consider the case of  
closed end boundary conditions (ring topology) and identify the occurrence of
partial phase synchronization via cluster formation.

Specifically, in the first part of this paper, we investigate the generic nature
of the phenomenon of global phase synchronization in an linear array, with free ends, of unidirectionally 
coupled (i) piecewise linear and (ii) threshold nonlinear time-delay systems in 
hyperchaotic regimes. In the second part, we report the phenomena of partial 
phase synchronizations in the array of above two systems but with closed end boundary conditions.
At first, we 
use the nonlinear transformation introduced in Senthilkumar {\it{et al.,}} [2006, 2008] 
to estimate the phase of all the systems in the array 
and to identify the existence of phase synchronization. Further, we confirm the existence
of global phase synchronization (GPS) and partial phase synchronization (PPS) from the 
original non-phase-coherent hyperchaotic attractors using two independent
approaches, namely recurrence quantification analysis \cite{romano05, marwan07} 
and the concept of localized sets \cite{pereira07}. 
In addition, we point out that the onset of GPS in the linear array with open end boundary conditions
takes place in the form of a sequential
synchronization (second order transition): For lower values of coupling strength the phases of nearby 
systems get already entrained 
with the drive system in contrast to the far away systems, while the
other non-synchronized systems display clusters of phase 
synchronized states among themselves before they become 
synchronized with the large cluster in the sequence as the strength of coupling increases to form the GPS.

On the other hand, if we consider an array with closed end boundary conditions
(ring topology), partial phase synchronization occurs with the formation of 
different groups of phase synchronized clusters.
The oscillators in the array self-organize to form groups of phase
synchronized clusters where each cluster is in perfect phase synchrony as
a function of the coupling strength. Such a clustering is
considered to be particularly significant in biological systems 
\cite{kaneko90, sherman94, strogatz93}, in electro-chemical oscillators \cite{kiss02}, etc. 
Recently, cluster synchronization in an array of three chaotic lasers without 
delay was reported \cite{terry99} as well. 
In our case, as the coupling strength increases, every individual system
starts to drive the nearest system and the systems with small differences in their phases
organize themselves to form small groups of clusters leaving the other oscillators 
with large phase differences to evolve independently. Further increase in the coupling 
strength results in an increase in the sizes of the clusters due to the locking 
of the phases of the nearby oscillators of the clusters for appropriate values of the coupling strength
and finally ending up with large groups of clusters resulting in PPS (first order transition).

The paper is organized as follows: In Secs.~\ref{sec:level2} and \ref{sec:level3},
we will describe briefly the coupling configuration 
and the occurrence of global phase synchronization in an array of two
different piecewise linear time-delay systems with open end boundary conditions. 
In Secs.~\ref{sec:level4} and \ref{sec:level5} we consider 
the linear array of two piecewise linear time-delay systems with closed end boundary 
conditions (ring topology) and discuss the occurrence of partial phase synchronization 
and finally we summarize our results in Sec.~\ref{sec:level6}. 

\begin{figure}
\centering
\includegraphics[width=0.5\columnwidth]{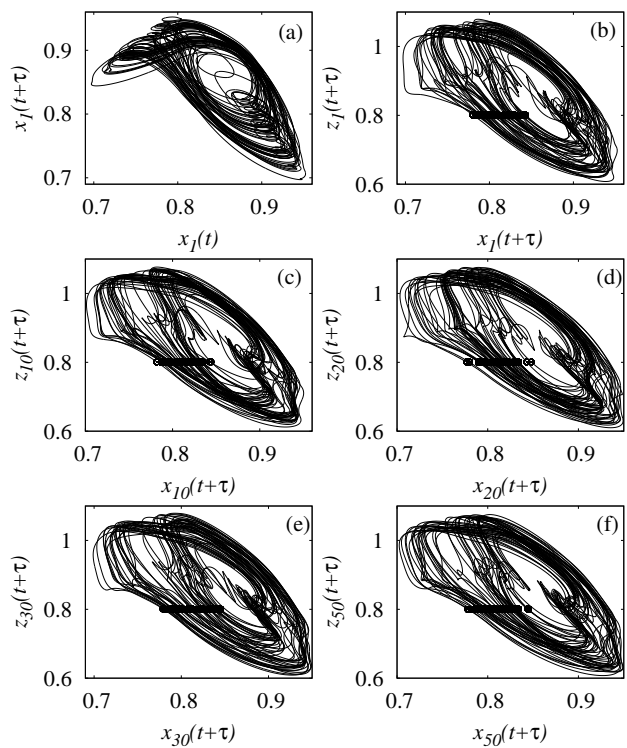}
\caption{\label{fig1} Attractors of the piecewise linear system. (a) non-phase-coherent hyperchaotic attractor of the drive 
system (\ref{eq_1}a). (b) Transformed attractor of the drive system. 
(c-f) Transformed attractors of some of the uniformly selected response systems 
($i = 10,20,30,50$) in the new state space ($x_i(t+\tau),z_i(t+\tau)$).}
\end{figure}
\section{\label{measures}Estimates of phase}
Identifying Chaotic phase synchronization is a nontrivial problem in time-delay systems
exhibiting complicated hyperchaotic attractors. For this purpose one requires 
appropriate measures. In this section, we describe briefly the various measures which 
we used from three different frameworks in this paper to estimate the phase of time-delay systems thereby facilitating
the characterization of phase synchronization between the coupled time-delay
systems. 
\subsection{Transformation of the original non-phase-coherent attractor}
Time-delay systems usually exhibit a highly complicated 
non-phase-coherent chaotic/hyperchaotic attractor with their flows having more 
than a single center of rotation. 
Such an attractor does not yield a monotonically increasing  phase on 
estimating it using the conventional techniques~\cite{senthilkumar06, senthilkumar08}.
For this purpose, we have introduced a nonlinear transformation to 
recast the original non-phase-coherent attractor into a smeared limit cycle-like attractor
with a single center of rotation. Our transformation is obtained by defining
a new state variable \cite{senthilkumar06, senthilkumar08},
\begin{equation}
z(t+\tau) = x(t)x(t+\hat{\tau})/x(t+\tau),
\label{eq_2}
\end{equation}
where $\hat{\tau}$ is the optimal value of time-delay to be chosen in order
to avoid any additional center of rotation. Then, the projected trajectory in
the transformed state space ($x(t+\tau),z(t+\tau)$) will resemble 
that of a smeared limit cycle-like attractor. Such an approach facilitates the estimation of
phase from the conventional techniques.  

\subsection{\label{rqa}Recurrence quantification analysis}
Several measures of complexity quantifying the small scale structures in 
recurrence plots have been proposed and the corresponding description is known as recurrence
quantification analysis (RQA)~\cite{marwan07}. In addition to the other 
advantages of the RQA such as its application to short experimental 
data, nonstationary and noisy data, we find that it can also be
applied directly to highly complicated non-phase-coherent attractors of the time-delay
systems in characterizing the synchronization transitions, in particular 
phase synchronization (PS) transitions. 
Among the available recurrence quantification measures, 
we use  the Correlation of Probability of 
Recurrence (CPR) and the generalized autocorrelation function $P(t)$, which are
estimated from the original non-phase-coherent hyperchaotic attractors,
to confirm the existence of GPS in the array of coupled time-delay systems,
both qualitatively and quantitatively.

A criterion to quantify phase synchronization between two systems
is the Correlation of Probability of Recurrence (CPR) defined as
\begin{align}
CPR=\langle \bar{P_1}(t)\bar{P_2}(t)\rangle/\sigma_1\sigma_2,
\label{cpr}
\end{align}
where $P(t)$ is the recurrence-based generalized autocorrelation function defined as
\begin{equation}
P(t)=\frac{1}{N-t} \sum_{i=1}^{N-t} \Theta(\epsilon-||X_i-X_{i+t}|| ),
\end{equation}
where $\Theta$ is the Heaviside function, $X_i$ is the $i^{th}$ data point of
the system $X$, $\epsilon$ is a predefined threshold, $|| . ||$ is the Euclidean 
norm, and $N$ is the number of data points, $\bar{P}_{1,2}$ means that the 
mean value has been subtracted and
$\sigma_{1,2}$ are the standard deviations of $P_1(t)$ and $P_2(t)$,
respectively.  Looking at the coincidence of the positions of the maxima of 
$P(t)$ of the coupled systems, one can qualitatively identify PS~\cite{romano05, marwan07}. 
If both  systems are in CPS, the probability of recurrence is
maximal at the same time $t$ and CPR $\approx 1.0$. If they are not in CPS,
the maxima do not occur simultaneously and hence one can expect a drift in both
probabilities of recurrences resulting in low values of CPR.  

\subsection{\label{ls} The concept of localized sets}
Another measure that we have employed is the framework of
localized sets~\cite{pereira07}.
The concept of localized sets opens up the possibility of an easy 
and an efficient way to detect CPS especially
in complicated non-phase-coherent attractors of time-delay systems. The main ingredient
of this technique is to define an event in one of the 
systems and then observe the other during the event, which
defines a set $D$. Depending upon the property 
of this set $D$, one can state whether PS exists or not. 
The set is spread over the entire attractor for asynchronous systems,
whereas it is localized on the attractor if the coupled systems are mutually
phase-locked or in phase synchronous state.

We use the above three different frameworks to confirm the
existence of GPS in an array of piecewise linear time-delay systems in the
following.

\section{\label{sec:level2}GPS in a Linear Array of Time-delay Systems with piecewise linearity}
Recently, we have reported the dynamical organization of an array of Mackey-Glass time-delay systems
in chaotic regime to form global phase synchronization (GPS) via sequential 
synchronization~\cite{suresh10}, as 
mentioned in the introduction. In this section, we intend to examine the generic nature of 
the results in Ref.~\cite{suresh10}, by investigating the emergence of 
GPS in an array of a particular type of piecewise linear time-delay systems from the
local sequential synchronization as 
the coupling strength is increased from zero to that of the hyperchaotic regimes. 
It is worth to emphasize that the hyperchaotic attractors of the piecewise linear
time-delay systems are characterized by
much more complex topological properties than the chaotic attractor of the
Mackey-Glass time-delay system studied in ~\cite{suresh10} and so is of considerable significance 
in estimating the onset of GPS through sequential synchronization.

\subsection{\label{sec:level2a}Linear array of piecewise linear time-delay systems}
Now, we consider an array of unidirectionally coupled piecewise linear 
time-delay systems with open end boundary conditions and with parameter mismatches.
Linear stability and bifurcation analysis of this system has been
well studied in Ref.~\cite{lakshmanan10,dvsijbc2005}. 
Many kind of synchronizations and their transitions have been 
reported in the coupled piecewise linear time-delay systems~\cite{senthilkumar05, senthilkumar07, lakshmanan10}. 
The array of unidirectionally coupled piecewise linear first-order delay differential equations
is represented as 
\begin{subequations}
\begin{eqnarray}
\dot{x}_1(t)&=&-\beta x_1(t)+ \alpha_{1} f(x_{1}(t-\tau)),  \\
\dot{x}_i(t)&=&-\beta x_i(t)+ \alpha_{i} f(x_{i}(t-\tau))+ \varepsilon(x_{i-1}(t)-x_{i}(t)), \quad i = 2, 3,\cdots, N,
\end{eqnarray}
\label{eq_1}
\end{subequations}
where $\alpha, \beta$ are the system parameters, $\tau$ is the time-delay and $\varepsilon$
is the coupling strength. We use the open end boundary conditions
$x_{0}=x_{1}$ and $x_{N}=x_{N-1}$. The nonlinear function $f(x)$
is chosen to be a piecewise linear function defined as
\begin{eqnarray}
f(x)=
\left\{
\begin{array}{cc}
0, &  x \leq -4/3  \\
            -1.5x-2,&  -4/3 < x \leq -0.8 \\
            x,&  -0.8 < x \leq 0.8 \\
            -1.5x+2,&  -0.8 < x \leq 4/3 \\
0, &  x > 4/3.
         \end{array} \right.
\end{eqnarray}
\begin{figure}
\centering
\includegraphics[width=0.7\columnwidth]{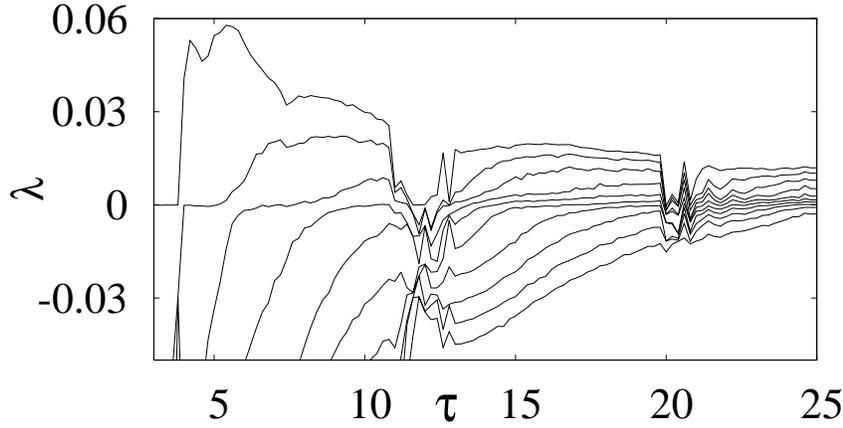}
\caption{\label{fig1b} Plot of the first ten maximal Lyapunov exponents $\lambda_{max}$
of the piecewise linear time-delay system (\ref{eq_1}a) for the parameter
values $\beta=1.0$, $\alpha_{1}=1.2, \tau \in(3,25)$.}
\end{figure}

The system parameters are fixed as follows: $\beta = 1.0$, $\alpha_1=1.2$, 
$\tau = 15.0$. The values of the nonlinear parameter $\alpha_i$ of the 
response systems in the array are chosen randomly in the range  
$\alpha_{i} \in (1.19,1.23)$, so that all the subsystems are effectively 
nonidentical due to the parameter mismatch.  The original non-phase-coherent 
hyperchaotic attractor of the drive in the $(x(t),x(t-\tau))$ state space is
illustrated in Fig.~\ref{fig1}(a). The corresponding transformed attractor,
effected using the transformation (\ref{eq_2}), in the new state space 
($x(t+\tau),z(t+\tau)$) is depicted in Fig.~\ref{fig1} (b),
which now looks like a smeared limit cycle-like attractor with a single 
center of rotation. The optimal value of
the $\hat{\tau}$ in (\ref{eq_2}) for the above piecewise linear time-delay
systems is found to be 1.6~\cite{senthilkumar06}. The same transformation
has also been performed for the non-phase-coherent chaotic attractor of 
the Mackey-Glass time-delay system with
an appropriate $\hat{\tau}$ in Ref.~\cite{suresh10}.
The hyperchaotic nature of the attractor in Figs.~\ref{fig1} for the above
parameter values is confirmed from three
positive Lyapunov exponents at $\tau = 15.0$ in the spectrum of ten maximal Lyapunov
exponents of (\ref{eq_1}a) as a function of the time-delay parameter $\tau \in (3, 25)$ 
in Fig.~\ref{fig1b}.

\begin{figure*}
\centering
\includegraphics[width=1.0\columnwidth]{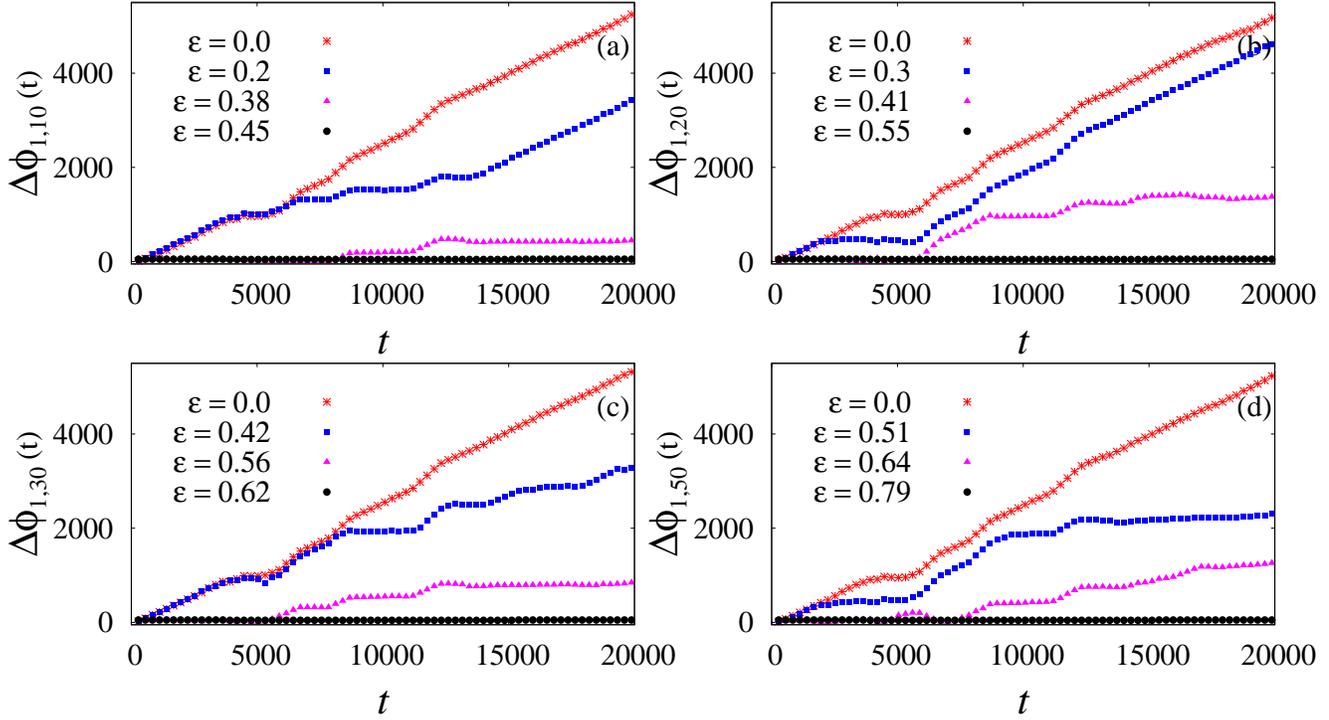}
\caption{\label{fig2} (a-d) Plots of the phase differences ($\Delta\phi_{1,i}$ = $\phi_{1}-\phi_{i}$) 
of some of the selected systems ($i$ = 10, 20, 30, 50) in the array
for different $\varepsilon$.}
\end{figure*}
\subsection{\label{sec:level2b}Identification of GPS from the transformed attractor of the 
piecewise linear time-delay system}
To confirm the existence of GPS in the array of piecewise linear time-delay systems, 
we have fixed size of the array as $N=50$. The transformed smeared limit cycle-like 
attractors with a single center of rotation of some of the uniformly selected subsystems 
($i$ = 10, 20, 30, 50) in the linear array are shown in Figs.~\ref{fig1} (c-f). 
We have used the conventional Poincar\'e section technique~\cite{pikovsky01, boccaletti02}
to estimate the instantaneous phases of all the oscillators after the transformation
of the attractors, which now yields monotonically increasing phase.
Open circles in Figs.~\ref{fig1}(b-f) indicate the Poincar\'e section.
\begin{figure}
\centering
\includegraphics[width=0.6\columnwidth]{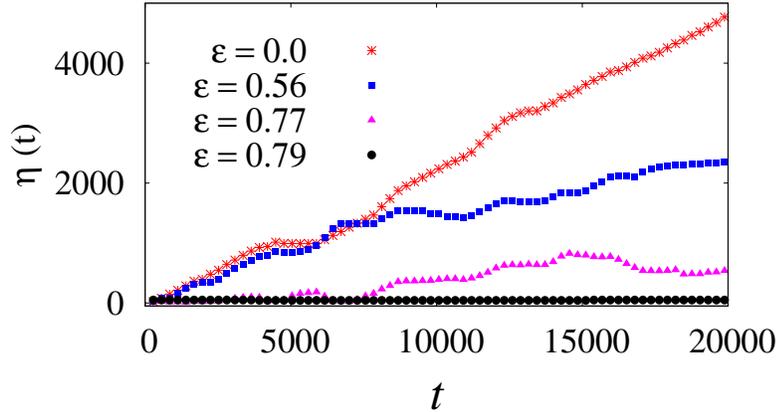}
\caption{\label{fig2a} The evolution of the average phase difference [$\eta(t)$]
for different values of the coupling strength $\varepsilon$.}
\end{figure}

In the absence of coupling ($\varepsilon = 0.0$) between the piecewise linear 
time-delay systems in the array,
the individual subsystems evolve independently in an asynchronous fashion which is indeed
confirmed from the monotonous increase in the phase differences, 
$\Delta\phi_{i} = \phi_{1}-\phi_{i}$, between the drive and the uniformly 
selected response systems ($i = 10,20,30,50$) indicated by asterisk symbols in 
Figs.~\ref{fig2}. Phase slips (marked by filled squares and triangles) 
in Figs.~\ref{fig2} for intermediate coupling strengths  indicate  that the
coupled systems are in their transition to phase synchronization with the
drive. For suitable values of $\varepsilon$, the response systems in the array are entrained
to the drive one sequentially, indicated by 
filled circles, as seen in Figs.~\ref{fig2}. To be more precise, 
it is clear from this figure that, the $10$th oscillator is synchronized first
at $\varepsilon = 0.45$, while the other systems away from it are still in their
transition to phase synchronization. Further increase in $\varepsilon$ to 
$\varepsilon=0.55$ results in synchronization of the $20$th oscillator with the drive
leaving the remaining far away oscillators from it in their transition state.  The
$30$th and $50$th oscillators are entrained to the drive at $\varepsilon = 0.62$ 
and $0.79$, respectively, illustrating the existence of GPS through the 
sequential synchronization of the subsystems locally in the array.
\begin{figure}
\centering
\includegraphics[width=0.7\columnwidth]{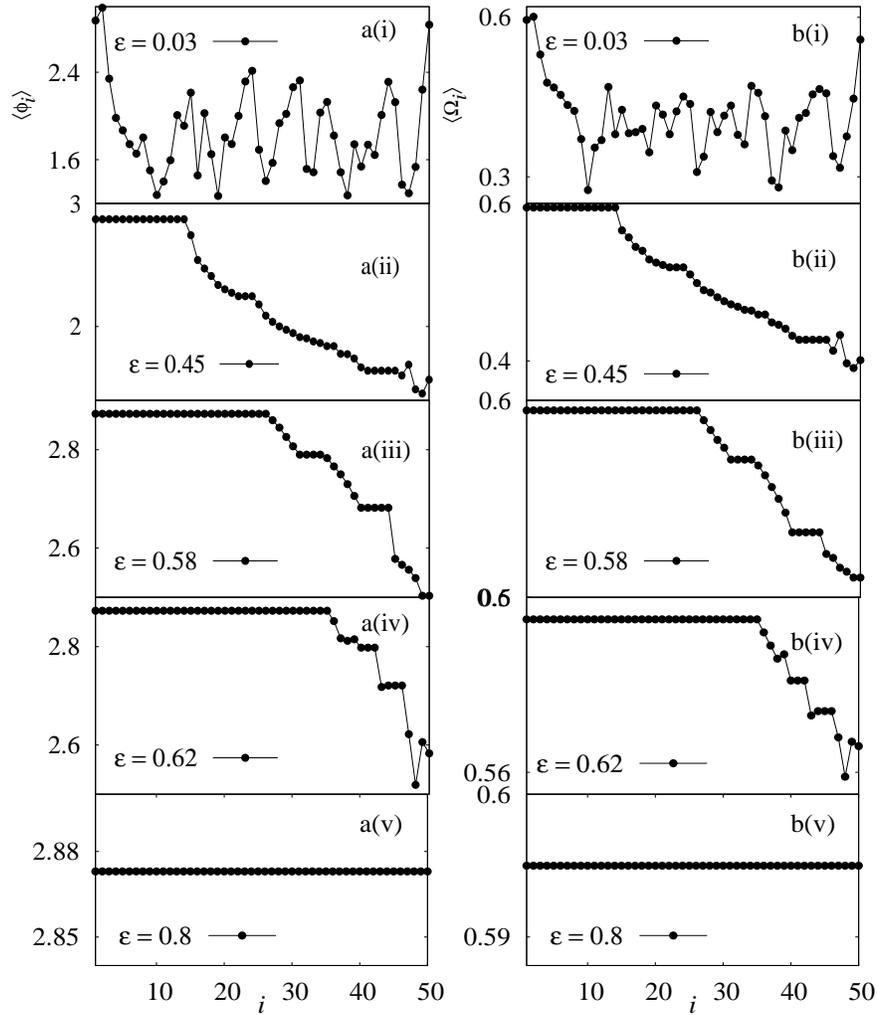}
\caption{\label{fig3} (a) Time averaged phase $\langle \phi_i \rangle$ given by (\ref{eq3a}) and (b) time averaged 
frequency $\langle \Omega_i \rangle$ given by (\ref{eq3a1}) of all the systems in the array (\ref{eq_1}) with open end boundary conditions plotted as a function of the system index $i$ for different values of the coupling strength $\varepsilon$.}
\end{figure}

The existence of GPS is also confirmed by the average phase
difference, $\eta(t)$, defined as
\begin{equation}
\eta(t) = \frac{1}{N-1} \sum_{j=2}^{N} (\phi_{1}-\phi_{j}).
\label{eq_3}
\end{equation}
A global measure to characterize the existence of GPS is the average phase difference ($\eta(t)$),
which is shown in Fig.~\ref{fig2a} for different values of $\varepsilon$. 
The monotonous increase in $\eta(t)$ for $\varepsilon=0.0$, 
correspond to the independent evolution of all the subsystems in the array. 
Phases of the nearby oscillators are mutually locked sequentially as $\varepsilon$ is 
gradually increased, which is indicated by the low degrees of $\eta(t)$ 
for $\varepsilon=0.56$ and $0.77$ in  Fig.~\ref{fig2a}.
Finally, $\eta(t)\approx 0$ at $\varepsilon=0.79$ corroborating the existence of GPS
in the array.
\begin{figure}
\centering
\includegraphics[width=0.8\columnwidth]{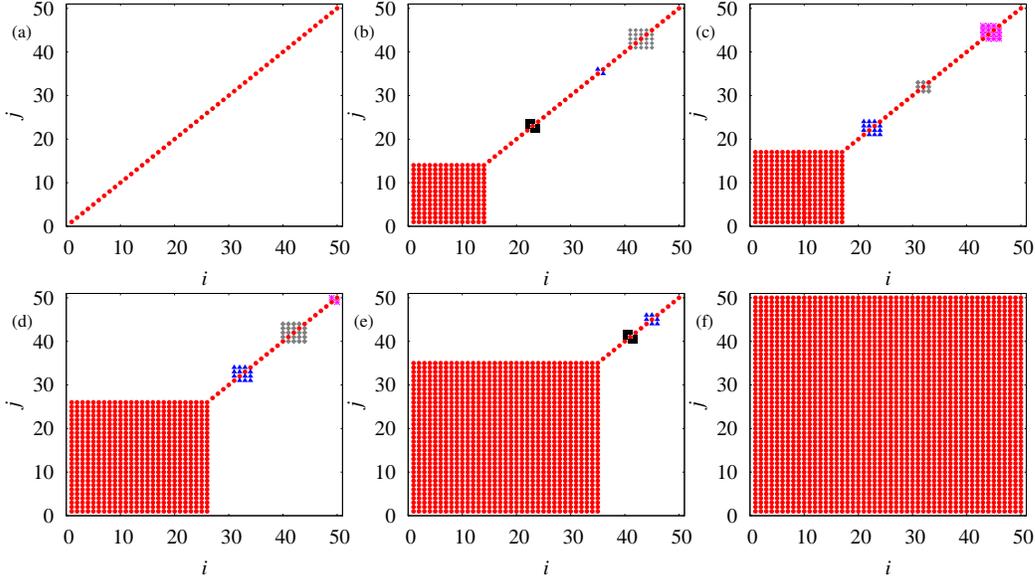}
\caption{\label{fig5} Oscillator index vs oscillator index snap shots indicating the
sequential phase synchronization and the dynamical
organization of cluster states for different values of coupling strength. 
The different shapes (colors) indicate that the corresponding nodes are phase 
synchronized. (a) non-phase synchronized case for $\varepsilon$ = 0.0 (b) First fourteen oscillators 
in Eq. (\ref{eq_1}b) are phase synchronized with the drive system for $\varepsilon$ = 0.45. (c), 
(d) and (e) Sequential phase synchronization and the formation 
of small cluster states for $\varepsilon$ = 0.51, 0.58 and 0.62, respectively, and 
(f) global phase synchronization for $\varepsilon$ = 0.8.}
\end{figure}

The nature of dynamical organization of local oscillators to form
GPS can be better understood by examining the time average phase and the
time average frequency as a function of the oscillator index for different values of 
the coupling strength. For this purpose, we define the time averaged phase ($\langle \phi_i \rangle$) as
\begin{equation}
\big\langle \phi_{i}(t)\big\rangle=\Big\langle2\pi k+2\pi \frac{t^{i}-t^{i}_{k}}{t^{i}_{k+1}-t^{i}_{k}}\Big\rangle_{t},~(t^{i}_{k}<t^{i}<t^{i}_{k+1}),
\label{eq3a}
\end{equation}
and the time averaged frequency as  ($\langle \Omega_i \rangle$) 
\begin{equation}
\big\langle \Omega_{i}(t)\big\rangle = \lim_{T \to \infty} \frac{1}{T} \int_{0}^{T} \dot\phi_{i}(t) dt,
\label{eq3a1}
\end{equation}
where $t_{k}$ is the time of the $k^{th}$ crossing of the flow with the  Poincar\'e 
section of the $i^{th}$ attractor and $\langle...\rangle_{t}$ denotes time average. 

Figures~\ref{fig3}(a) and~\ref{fig3}(b) depict 
the average phase and the average frequency, respectively, for different $\varepsilon$
as a function of the oscillator index $i$. The random distribution of
$\big\langle \phi_{i}(t)\big\rangle$ (Fig.~\ref{fig3}(a)(i)) and 
$\big\langle \Omega_{i}(t)\big\rangle$ (Fig.~\ref{fig3}(b)(i))
for $\varepsilon=0.0$ corresponds to the large difference in the 
initial frequency of the uncoupled oscillators due to the nonlinear parameter mismatch.
The scenario of sequential phase synchronization to reach GPS can also be
clearly visualized using the snap shots of the oscillators in the index vs index plot
as shown in  Fig.~\ref{fig5}. The diagonal line in Fig.~\ref{fig5}(a) for $\varepsilon=0.0$
clearly shows the asynchronous evolution of all the oscillators in the array.

The phase locking of the first $14$ oscillators with the drive, 
forming the main cluster, for $\varepsilon=0.45$ 
is shown in Figs.~\ref{fig3}(a)(ii) and ~\ref{fig3}(b)(ii),
while some of the other oscillators (oscillators with indices $22-24$, $35-36$ 
and $41-45$) away from it form  phase synchronized clusters among themselves. 
The main cluster and other small clusters of phase synchronized oscillators
are much more clearly visualized in Fig.~\ref{fig5}(b) for the same coupling strength,
in which oscillators with same frequency are assigned the same shape and color.
The average phase and the average frequency for $\varepsilon=0.58$ depicted in
Figs.~\ref{fig3}(a)(iii) and Fig.~\ref{fig3}(b)(iii) indicate that the 
first $26$ oscillators are synchronized with the drive with a few other small
synchronized clusters away from it. Similarly, the first $35$ oscillators are
synchronized with the drive for $\varepsilon=0.62$ (see Figs.~\ref{fig3}(a)(iv) 
and Fig.~\ref{fig3}(b)(iv)) along with other small clusters. 
Thus, it is clear that an
increase in the coupling strength results in desynchronization of some of
the existing clusters and formation of new clusters away from the main cluster. 
At the same time, nearby desynchronized oscillators from the main cluster becomes
synchronized sequentially with it thereby increasing its size,
contributing to the basic mechanism of
formation of GPS via sequential synchronization of local oscillators
as explained in detail in Ref.~\cite{suresh10}.
Finally for 
$\varepsilon=0.8$, all the oscillators become phase locked to attain GPS
as depicted in Figs.~\ref{fig3}(a)(v) and ~\ref{fig3}(b)(v), which remains
stable for further larger values of the coupling strength. 
For the above values of $\varepsilon$, the main
cluster and other small clusters are also clearly seen in the
index vs index plot in Figs.~\ref{fig5}. 

\begin{figure}
\centering
\includegraphics[width=0.6\columnwidth]{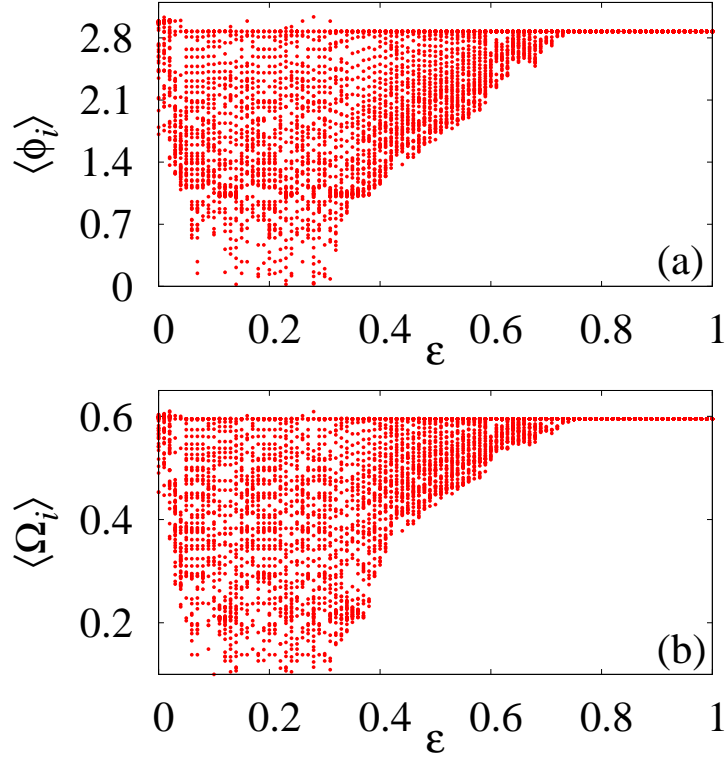}
\caption{\label{fig6} (a) Time averaged phase ($\langle \phi_i \rangle$) and 
(b) Time averaged frequency ($\langle \Omega_i \rangle$), $i = 1,2,\cdots, 50$ as a function of 
the coupling strength $\varepsilon \in (0,1.0)$. Here for each value of $\varepsilon$
we have plotted the average phase/frequency of all the $N$ = 50 oscillators which is shown 
by the filled circles.}
\end{figure}

Further, we have also plotted the 
average phase ($\langle \phi_i \rangle$) and the average frequency 
($\langle \Omega_i \rangle$) of all the $N$ oscillators as a function of 
$\varepsilon$ in Figs.~\ref{fig6} to get a global picture 
of the above dynamics. Broad distribution of $\langle \phi_i \rangle$ and
$\langle \Omega_i \rangle$ for small $\varepsilon$ correspond to asynchronous 
states while for $\varepsilon\ge 0.75$ both the average phase and  average frequency 
converge to a single value indicating the existence of GPS. However, for the
intermediate values of $\varepsilon$, the main cluster and other small
clusters are hardly visible in Figs.~\ref{fig6}. Nevertheless, these
clusters are clearly seen in Figs.~\ref{fig3} and~\ref{fig5}, where the
snapshots of the dynamical organization of all the $N$ oscillators in the
array are depicted for specific values of the coupling strength.

\begin{figure}
\centering
\includegraphics[width=0.6\columnwidth]{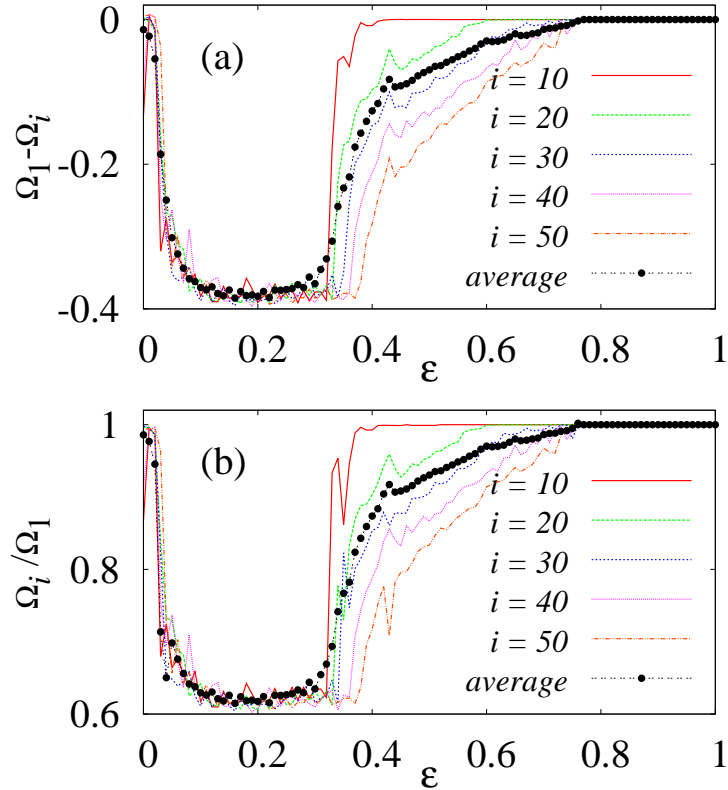}
\caption{\label{fig8} (a) The frequency difference ($\Delta \Omega_{1,i}, 
\quad i=10,20,30,40$ and $50$) and (b) the frequency ratio ($\Omega_i/\Omega_1, 
\quad i=10,20,30,40$ and $50$) of uniformly selected systems are plotted as a function of the coupling strength 
$\varepsilon \in (0,1.0)$. Each line corresponds to the difference/ratio between the 
response system and the drive system. The black filled circles indicate the average
frequency difference/ratio of all the ($N-1$) response systems from the drive system.}
\end{figure}

The frequency difference ($\Delta \Omega_{1,i}=\Omega_1-\Omega_i$) 
and the frequency ratio ($\Omega_i/\Omega_1$) for the oscillators with 
the index $i=10,20,30,40$ and $50$ along with their average (represented by black 
filled circles) of all the $(N-1)$ response oscillators as a function of the
$\varepsilon$ are depicted in Fig.~\ref{fig8}(a) and Fig.~\ref{fig8}(b), respectively.
It is also clearly seen from these figures that the nearby oscillators 
are synchronized sequentially as $\varepsilon$ is increased, that is, $i=10$ is
synchronized first and then $i=20$, and so on as $\varepsilon$ is increased. Finally, all the
$N=50$ oscillators in the array are phase synchronized for 
$\varepsilon\ge 0.75$ attributing to the emergence of GPS via local sequential 
synchronization.

\begin{figure}
\centering
\includegraphics[width=0.7\columnwidth]{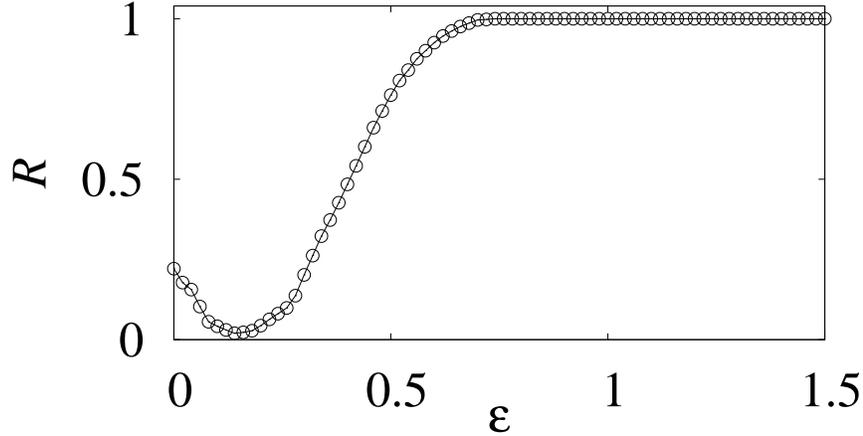}
\caption{\label{fig10} Plot of the phase order parameter ($R$) as a function of 
the coupling strength indicating global phase synchronization in the array 
of ($N$ = 50) coupled piecewise linear time-delay systems (\ref{eq_1}).}
\end{figure}

The well-known Kuramoto order parameter \cite{moreno04} can also be used for
quantifying the global phase synchronization in the array, which is defined as
\begin{equation}
R~e^{i\psi} = \Big\langle\Big|\frac{1}{N}\sum_{j=1}^{N} e^{i\phi_{j}(t)}\Big|\Big\rangle_t.
\label{eq_kop}
\end{equation}
Here $\phi_{j}(t)$ represents the instantaneous phase of the $j^{th}$ system, $\psi(t)$ is the 
average phase and $\langle...\rangle_t$ denotes a time average. When all the systems are in a phase 
synchronized state, the value of $R\approx 1.0$. Figure~\ref{fig10} shows that $R$ decreases at
first for small increase in $\varepsilon$ as observed in Figs.~\ref{fig8}(a) and \ref{fig8}(b)
accounting for the increased randomness in the phases of the coupled systems as
seen in the average phase and the average frequency of all the oscillators
in Figs.~\ref{fig6}. Further increase in $\varepsilon$ results in a gradual
increase in the value of $R$, reaching $R\approx 1.0$ for $\varepsilon>0.75$ confirming
the existence of GPS in the array of coupled piecewise linear time-delay systems.

It is also to be noted that the phenomenon remains qualitatively the 
same even when the total number of oscillators in the array is increased. 
We also wish to emphasize that the results remain qualitatively unaltered 
even for different sets of random values for the nonlinear parameters, 
$\alpha_i$, confirming the robustness of our results. 

\subsection{\label{sec:level2c}Confirmation of GPS from the original untransformed non-phase-coherent attractor}

Next, we use two different approaches, namely recurrence quantification analysis \cite{romano05, marwan07} and 
the concept of localized sets \cite{pereira07} without estimating the phase explicitly to prove the 
existence of GPS from the original nonphase coherent hyperchaotic attractors. 

\subsubsection{\label{sec:level2d}Recurrence quantification for GPS}
\begin{figure*}
\centering
\includegraphics[width=0.9\columnwidth]{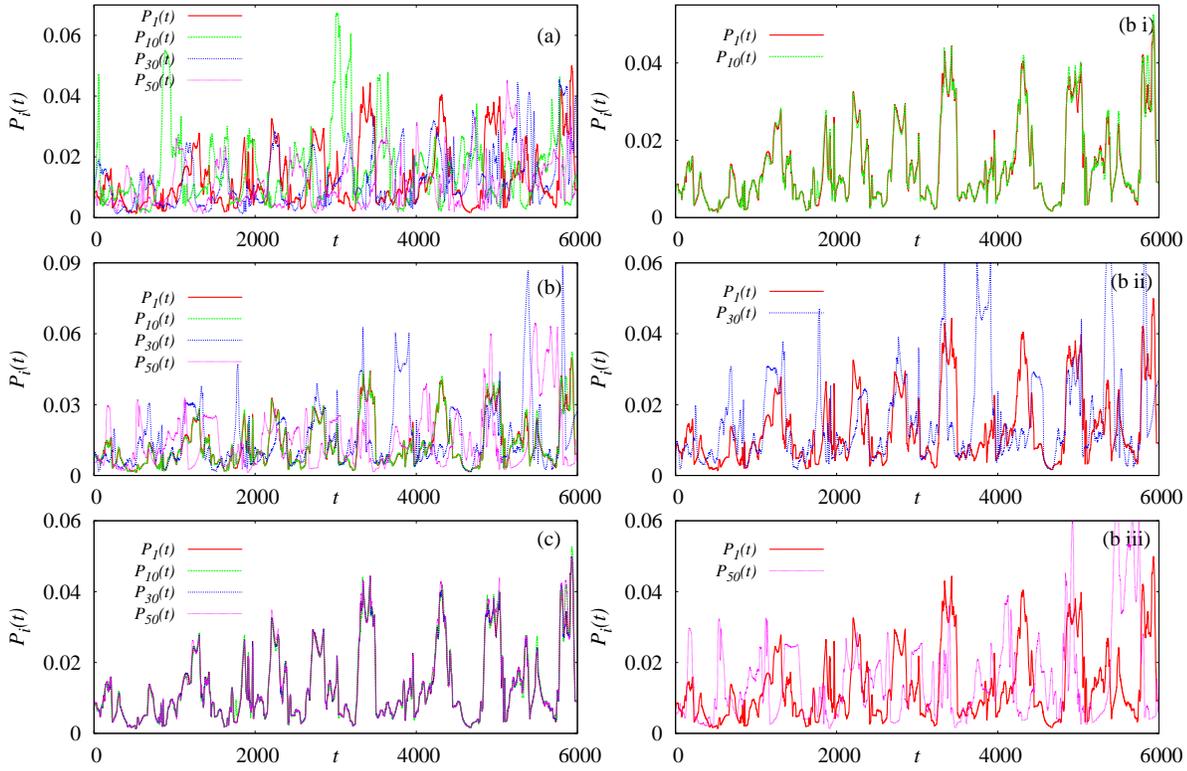}
\caption{\label{fig11} Plots of the Generalized autocorrelation functions of the 
drive $P_{1}(t)$ and some selected response systems ($i=10,30$,and, $50$) $P_{10}(t)$, 
$P_{30}(t)$, and $P_{50}(t)$ indicating (a) Non-phase-synchronization for $\varepsilon=0.0$,
(b) generalized autocorrelation functions  for $\varepsilon=0.45$
(bi) PS  between the systems $1$ and $10$, (bii) approximate PS between the systems $1$ and
$30$ and (biii) non PS between the systems $1$ and $50$, and (c) PS between all 
the systems ($i=1,10,30$, and $50$) for $\varepsilon=0.8$.}
\end{figure*}

Now, we use the recurrence quantification measures discussed
in Sec.~\ref{rqa} in characterizing the existence of GPS via
sequential synchronization. Figure~\ref{fig11} depicts the
generalized autocorrelation function of the drive $P_1(t)$ and that of
some of the response systems $P_{10}(t)$, $P_{30}(t)$, and $P_{50}(t)$
for different $\varepsilon$. For $\varepsilon=0.0$, none of the maxima of 
$P_i(t)$ coincide in Fig.~\ref{fig11}(a) indicating asynchronous state.
$P_i(t)$ for $i=1,10,30$ and $50$ are plotted in Figs.~\ref{fig11}(b) and
~\ref{fig11}(c) for $\varepsilon=0.45$ and $\varepsilon=0.8$, respectively.
In Fig.~\ref{fig11}(b),  the maxima of $P_1(t)$ and $P_{10}(t)$ are in perfect
agreement (see Fig.~\ref{fig11}(b)(i)) confirming the existence of phase synchronization 
between the oscillators with the index $i=1$ and $i=10$, which is in agreement
with the results of the earlier section. On the other hand, some of the 
maxima of $P_1(t)$ and $P_{30}(t)$ in Fig.~\ref{fig11}(b)(ii) are in agreement 
attributing to the transition to CPS among them and
none of the maxima of $P_1(t)$ and $P_{50}(t)$ in Fig.~\ref{fig11}(b)(ii) are in agreement
corresponding to independent evolution, which confirms the sequential synchronization. 
All the maxima of $P_i(t)$ are in complete
agreement in Fig.~\ref{fig11}(c) confirming the existence of GPS.
Also, the formation of
clusters by the other asynchronous oscillators in the array can be confirmed by
plotting their respective generalized autocorrelation functions, which will show that
all their maxima are in good agreement with each other, whereas there exists a drift 
between them and the maxima of the sequentially synchronized cluster as 
discussed in Ref.~\cite{suresh10}.

\begin{figure}
\centering
\includegraphics[width=0.7\columnwidth]{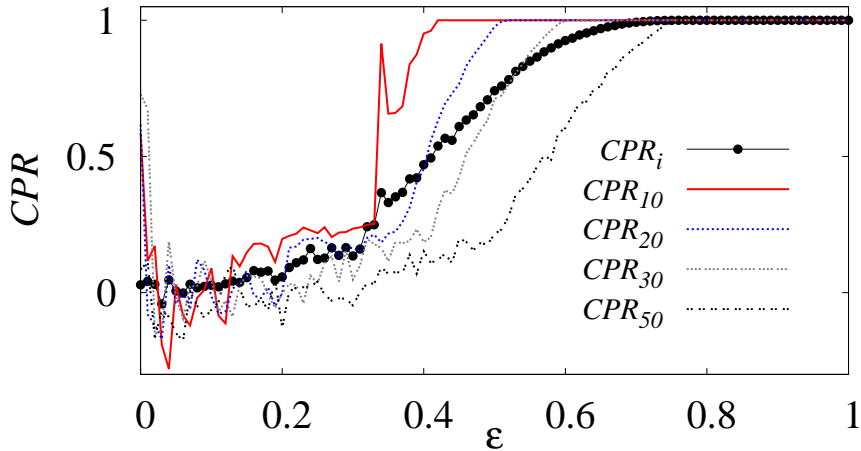}
\caption{\label{fig12} Plots of the index CPR as a function of the coupling strength $\varepsilon$. 
Different lines correspond to the CPR of different ($i=10,20,30$ and $50$) response 
systems with the drive system. The filled circles correspond to the mean value of the CPR 
of all the ($N-1$) piecewise linear systems in the array.}
\end{figure}

The global scenario of 
the existence of GPS via sequential phase synchronization can also 
confirmed using the index CPR in analogy with the nature of the Kuramoto order
parameter $R$, the average frequency difference and the average
frequency ratio. The index CPR of the response systems $i=10, 20, 30, 50$ with that 
of the drive is shown in Fig.~\ref{fig12} as a function of $\varepsilon$.  
It is clear that the nearby oscillators to the drive are synchronized first
as the coupling strength is gradually increased contributing to the local sequential
synchronization.
The results are exactly similar to that observed from the average frequency 
difference and the average frequency ratio in Figs.~\ref{fig8} indicating
the phenomenon of GPS via sequential phase synchronization. 
\begin{figure*}
\centering
\includegraphics[width=1.0\columnwidth]{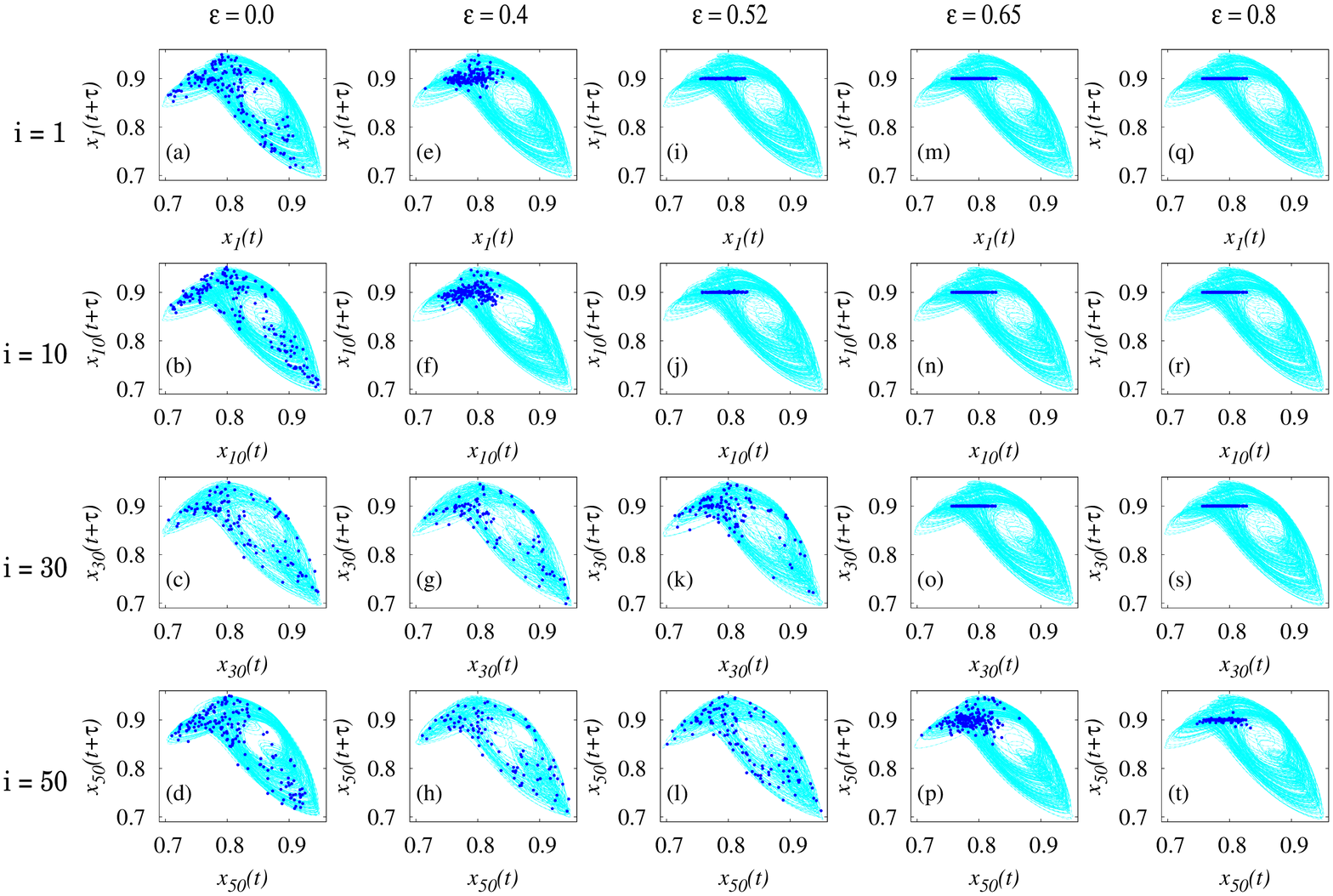}
\caption{\label{fig13} First row (a-q) corresponds to the attractors of the drive system ($i = 1$) 
and the rows (b-r, c-s, and d-t) correspond to the
attractors of some  selected response systems ($i = 10,30,50$). 
In (a-d) the sets (represented by the filled circles) 
are spread over the attractors and hence 
there is no CPS for the value of coupling strength $\varepsilon$ = 0.0. In (e-h) for 
$\varepsilon$ = 0.4 and in (i-l, m-p, q-t) the sets 
are localized confirming the existence of GPS in the array for $\varepsilon$ = 0.52, 0.65, and 0.8, respectively.}
\end{figure*}

\subsubsection{\label{sec:level2e}GPS using the concept of localized sets}
Next, we use the framework of localized sets described in Sec.~\ref{ls} 
to demonstrate the existence of GPS via sequential phase synchronization.
We defined the event as a Poincar\'e section  chosen at $x(t-\tau)=0.8$ 
and $x(t)<0.8$ on the attractor. 
The sets (indicated as filled circles) obtained by observing the drive 
system ($i = 1$) whenever the defined event occurs in the response system $i = 10$ 
along with the drive attractor are depicted in the first row in Fig.~\ref{fig13}.
Similarly, the sets obtained by observing the response systems ($i = 10, 30$ and $50$) 
whenever the event occurs in the drive system are depicted along with the
corresponding response attractors in the other subsequent rows.

The sets spread over the entire attractors in Figs.~\ref{fig13}(a-d) for 
$\varepsilon = 0.0$ confirm the asynchronous evolution of the
subsystems in the array. The sets localized in a large area of
the attractors in Figs.~\ref{fig13}(e-f) for $\varepsilon = 0.4$ indicate the transition
to CPS among the corresponding oscillators, whereas the other oscillators
away from it still evolve independently as indicated by the spread of
the sets over their attractors (see Figs.~\ref{fig13}(g-h)). The more localized
sets in  Figs.~\ref{fig13}(i-j) for $\varepsilon = 0.52$ indicate that
the respective oscillators are in complete phase synchrony while the
other systems away from it are in asynchronization with it as confirmed by
the spread of the sets in Figs.~\ref{fig13}(k-l). For $\varepsilon = 0.65$,
the sets are bounded to smaller regions of the attractors in Figs.~\ref{fig13}(m-o)
attributing to the existence of CPS between the drive and subsystems $10$ and $30$. 
The other subsystem  with index $i=50$ is in its transition state as indicated
by the localized set but in a large region of the attractor as seen in Fig.~\ref{fig13}(p). 
Thus it is clear from these figures that the oscillators away from the drive
system are synchronized sequentially as the coupling strength is increased gradually. 
Finally, for $\varepsilon = 0.8$, the sets are localized to a narrow region on
the attractors (Figs.~\ref{fig13}(q-t)) of all the subsystems thereby confirming the existence of GPS.

\begin{figure}
\centering
\includegraphics[width=0.7\columnwidth]{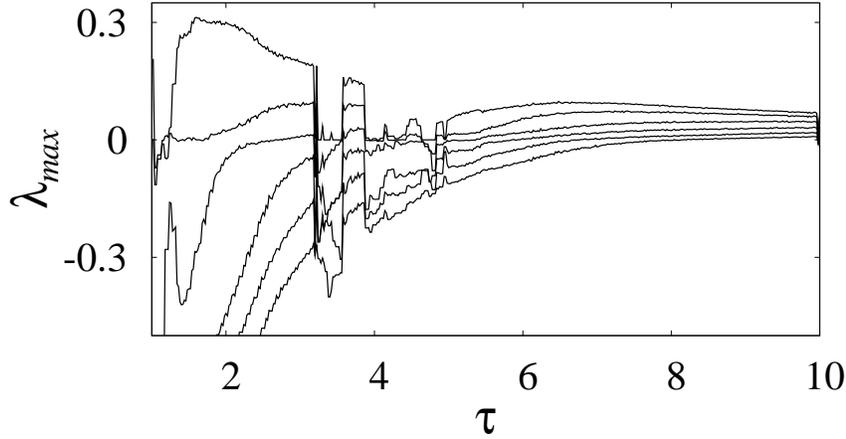}
\caption{\label{fig13a} The first six maximal Lyapunov exponents $\lambda_{max}$
of the threshold piecewise linear time-delay system (Eq.~(\ref{eq_3pl})) for the parameter
values $\alpha=1.2, \beta=1.0, \tau \in(1,10)$ in the absence of the coupling $\varepsilon$.}
\end{figure}

\section{\label{sec:level3}Global Phase Synchronization in an Array of Piecewise Linear Systems with Threshold Nonlinear Function}
In this section, we demonstrate the existence of the GPS in another piecewise linear 
time-delay system with a threshold nonlinear function. This system has been studied recently
for its hyperchaotic nature even for small values of time-delay [see for details, Lakshmanan \& Senthilkumar, 2010]
 and has also been experimentally realized using analog electronic circuits \cite{srinivasan11}. 
Very recently, the chaotic phase
synchronization has been experimentally confirmed in this piecewise linear time-delay system along with
numerical simulation \cite{dvs10} and various types of synchronization transitions 
have been demonstrated both experimentally and numerically in this system \cite{srinivasan11}. 
In this paper, we consider a linear array of piecewise linear systems as in 
Eq. (\ref{eq_1}) and the nonlinear function $f(x)$ is now chosen to be a piecewise linear 
function with a threshold nonlinearity,
\begin{equation}
\label{eq_3pl}
f(x)=AF^{*}-Bx.
\end{equation}
Here
\begin{eqnarray}
F^{*}=
\left\{
\begin{array}{cc}
-x^{*},&  -x < x^{*}  \\
            x,&  -x \leq x \leq x^{*} \\
            x^{*},&  x > x^{*} \\ 
         \end{array} \right.
\label{eq_4}
\end{eqnarray} 
where $x^{*}$ is a controllable threshold value and $A$ and $B$ are positive parameters.
This function $f(x)$ employs a threshold controller for flexibility. It effectively implements
a piecewise linear function. The control of this piecewise linear function facilitates
controlling the shape of the attractors. Even for a small delay value, this system 
exhibits hyperchaos and can produce multi-scroll chaotic attractors by just introducing more
threshold values. In our analysis, we chose $x^{*}$ = 0.7, $A$ = 5.2, $B$ = 3.5,
$\alpha_{1}$ = 1.2, $\beta$ = 1.0, $\tau$ = 6.0 and the nonlinear parameters $\alpha_{i}$,
of the response systems in the array are chosen randomly in the range $\alpha_{i} \in (1.18, 1.24)$.
Note that for this set of parameter values a single uncoupled system exhibits a highly 
complicated hyperchaotic attractor with three positive Lyapunov exponents. The first six
largest Lyapunov exponents of the uncoupled system are shown in Fig.~\ref{fig13a} as a 
function of time-delay $\tau \in (1,10)$. 

We have not yet succeeded in generalizing the
nonlinear transformation (Eq. (\ref{eq_2})) to capture the phase of the non-phase-coherent
hyperchaotic attractor of the above systems due to the multiscroll attractor. 
However, we find that the recurrence-based indices serve as excellent quantifiers in identifying the transition
from non-synchronized to phase synchronized state both quantitatively and qualitatively. 
We have also characterized the occurrence of GPS using the concept of localized sets.

\subsection{\label{sec:level3b}GPS using recurrence analysis}
\begin{figure*}
\centering
\includegraphics[width=0.9\columnwidth]{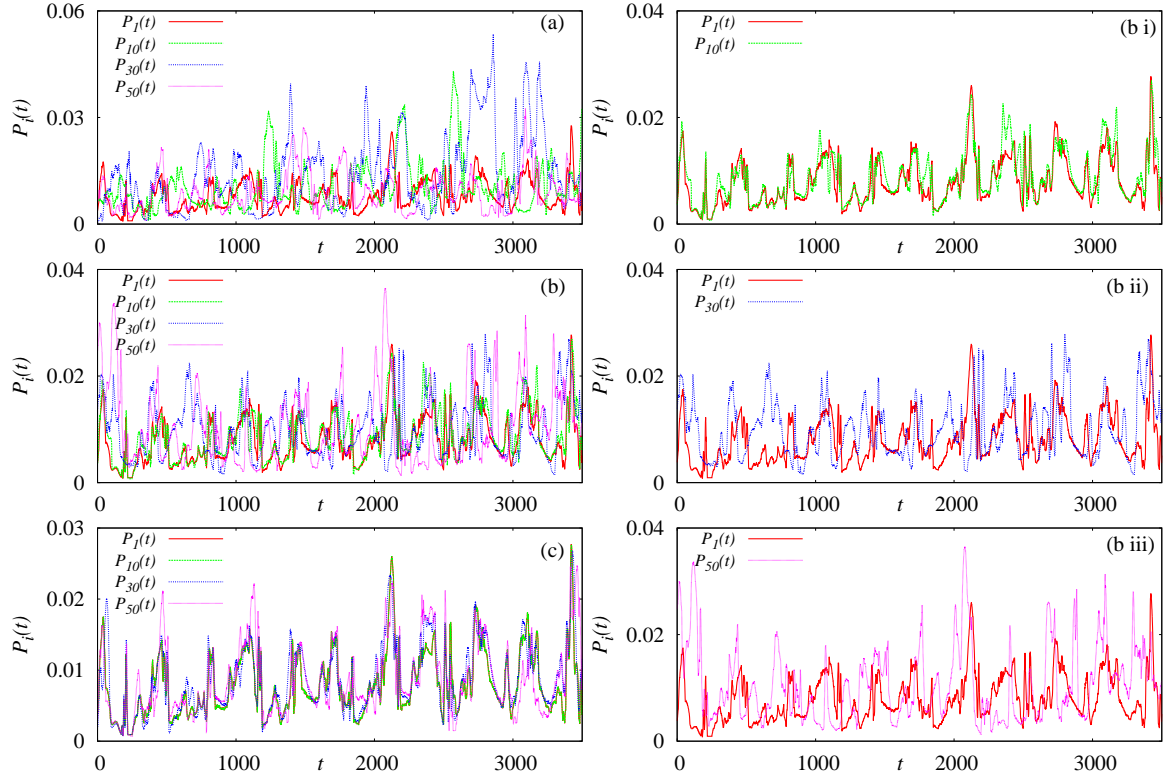}
\caption{\label{fig13b} Generalized autocorrelation functions of the 
drive $P_{1}(t)$ and some of the selected response systems ($i=10,30$,and, $50$) $P_{10}(t)$, 
$P_{30}(t)$, and $P_{50}(t)$ indicating (a) Non-phase-synchronization for $\varepsilon=0.0$,
(b) generalized autocorrelation functions  for $\varepsilon=0.8$
(bi) PS  between the systems $1$ and $10$, (bii) approximate PS between the systems $1$ and
$30$ and (biii) non-PS between the systems $1$ and $50$, and (c) PS between all 
the selected systems ($i=1,10,30$, and $50$) for $\varepsilon=1.2$.}
\end{figure*}

We have calculated the CPR and the generalized autocorrelation function $P(t)$ to confirm the existence 
of the GPS in the array. The generalized autocorrelation function of the drive $P_1(t)$ and that of
some response systems ($i=10,30$, and $50$), $P_{10}(t)$, $P_{30}(t)$, and $P_{50}(t)$,
are depicted in  Fig.~\ref{fig13b} for different values of the coupling strength.
In the absence of coupling ($\varepsilon=0.0$), all the systems evolve independently and
hence the maxima of their respective generalized autocorrelation functions do not occur
simultaneously as shown in Fig.~\ref{fig13b}(a). On increasing the coupling strength,
the oscillators with a lower value of index in the array become synchronized first resulting in 
sequential phase synchronization and this can also be identified from the
generalized autocorrelation functions of the response systems in the array.
For instance, $P_{10}(t)$, $P_{30}(t)$, and $P_{50}(t)$ are shown along with $P_{1}(t)$ in 
Fig.~\ref{fig13b}(b) for $\varepsilon=0.8$. It is clear
from this figure that the maxima of the drive $P_{1}(t)$ and those of the
response $P_{10}(t)$ are in complete agreement with each other [Fig.~\ref{fig13b}(b)(i)] indicating the
existence of PS between them. On the other hand, only some of the maxima of the
response system $P_{30}(t)$ are in coincidence with those of the drive [Fig.~\ref{fig13b}(b)(ii)] 
illustrating that the response system $i=30$ is in transition to PS,
whereas the maxima of the response system $P_{50}(t)$ do not coincide with
those of the drive [Fig.~\ref{fig13b}(b)(iii)] indicating that the response system $i=50$ is in an asynchronous 
state for the same value of $\varepsilon$. For $\varepsilon=1.2$, almost all of the positions of 
the peaks of the generalized autocorrelation functions $P_{1}(t)$, $P_{10}(t)$, 
$P_{30}(t)$, and $P_{50}(t)$ are in agreement with each other as illustrated 
in Fig.~\ref{fig13b}(c) confirming the existence of GPS via sequential phase
synchronization. It is also to be noted that the magnitudes 
of the peaks of all the oscillators have generally different values and 
these differences in the heights of the peaks indicate that there is no 
correlation in the amplitudes of the coupled systems.
\begin{figure}
\centering
\includegraphics[width=0.8\columnwidth]{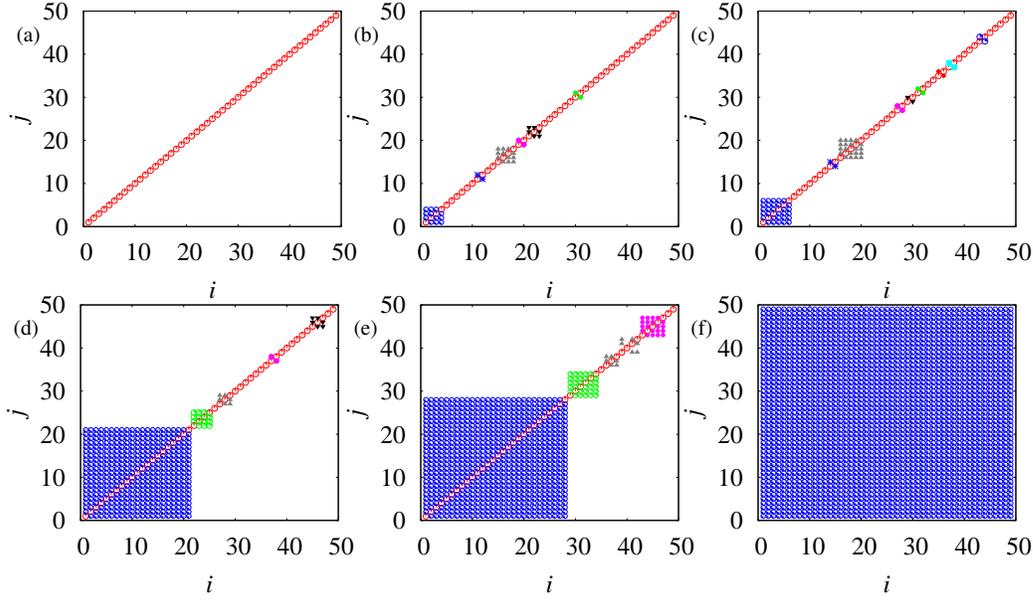}
\caption{\label{fig13c} Snap shots of the node vs node diagrams 
indicating the sequential phase synchronization and the
organization of cluster states for different values of coupling strength for system (\ref{eq_1})
with the threshold nonlinearity (\ref{eq_3pl}). 
(a) non-phase synchronized case for $\varepsilon$ = 0.0 (b) First four oscillators 
are phase synchronized with the drive system for $\varepsilon$ = 0.65. (c), 
(d) and (e) Sequential phase synchronization and the formation 
of small cluster states for $\varepsilon$ = 0.7, 0.94 and 1.13, respectively, and 
(f) global phase synchronization for $\varepsilon$ = 1.25.}
\end{figure}

The dynamical organization of GPS via sequential synchronization and the clustering
can be visualized clearly by plotting some snap shots of the oscillators in the index vs index
plots as shown in Fig.~\ref{fig13c}. The oscillators that evolve with identical phase are assigned
with identical shapes (colors). The diagonal line in Fig.~\ref{fig13c}(a) for $\varepsilon=0.0$
correspond to the oscillator index $i=j$ and the oscillators evolve independently. Figure \ref{fig13c}(b)
indicates that the first four oscillators in the array are synchronized with the drive for 
$\varepsilon=0.65$ along with the five small separate clusters. In Fig.~\ref{fig13c}(c) for 
$\varepsilon=0.7$ the first six oscillators form a synchronized cluster along with eight other
small clusters. Similar small clusters are shown in Fig.~\ref{fig13c}(d) and (e) for $\varepsilon=0.94$
and $\varepsilon=1.13$, respectively, in addition to the single large cluster formed by sequential 
phase synchronization. Finally GPS of all the systems in the array is illustrated in Fig.~\ref{fig13c}(f)
for $\varepsilon=1.25$.

\begin{figure}
\centering
\includegraphics[width=0.7\columnwidth]{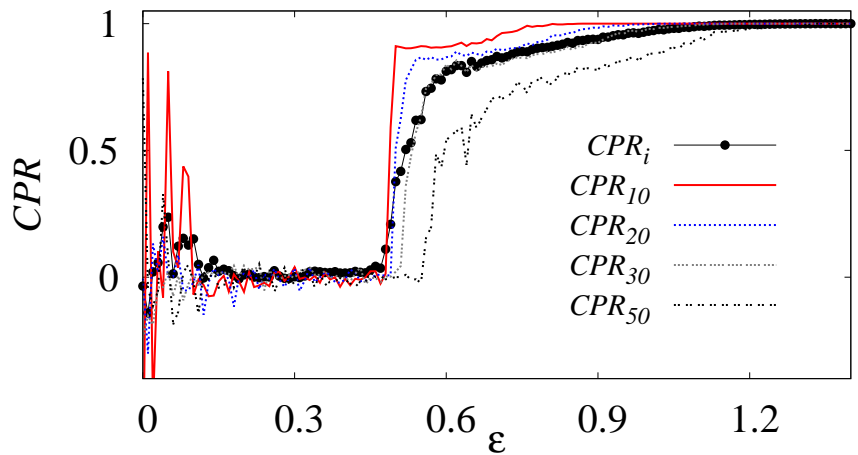}
\caption{\label{fig14} The index CPR as a function of the coupling strength $\varepsilon$. 
Different lines correspond to the CPR of different ($i=10,20,30,$ and $50$) response 
systems with the drive system. The filled circles correspond to the mean value of the CPR 
of all the ($N-1$) piecewise linear systems in the array.}
\end{figure}

The existence of GPS via sequential phase synchronization is  also quantified 
using the index CPR (\ref{cpr}) of the response systems with the drive as shown in Fig.~\ref{fig14}. 
The different lines correspond to the index of the oscillators ($i$ = 10, 20, 30, 50) in the
array. It is evident from the figure that the oscillators with increasing index attain
the value of unity in a sequence as a function of the coupling strength and finally
for $\varepsilon>1.15$ the CPR of all the response systems with the drive reaches
unity confirming that all the coupled oscillators are in GPS.  
The mean value of CPR of all the response systems in the array, shown as filled circles, 
also confirms the existence of GPS for $\varepsilon>1.15$.
\begin{figure*}
\centering
\includegraphics[width=1.0\columnwidth]{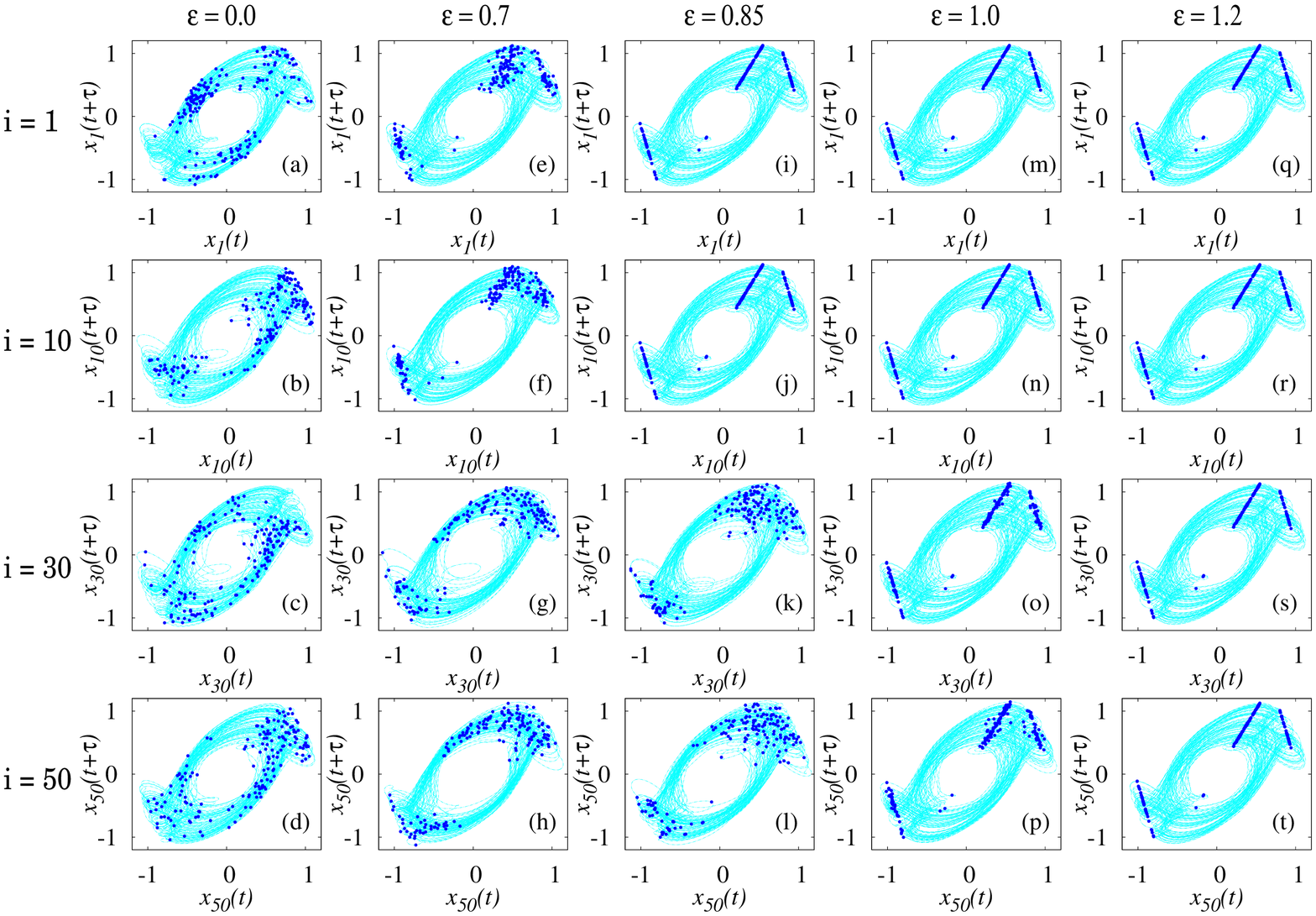}
\caption{\label{fig15} First row (a-q) corresponds to the attractors of the drive system ($i = 1$) 
and the rows (b-r, c-s, and d-t) correspond to the
attractors of some selected response systems ($i = 10,30,50$). 
In (a-d) the sets (represented by the filled circles) 
are spread over the attractors and hence 
there is no CPS for the value of coupling strength $\varepsilon$ = 0.0. In (e-h) for 
$\varepsilon$ = 0.7 and in (i-l, m-p, q-t) the sets 
are localized confirming the existence of GPS in the array for $\varepsilon$ = 0.85, 1.0, and 1.2, respectively.}
\end{figure*}

\subsection{\label{sec:level3c}GPS using the concept of localized sets}
We have also confirmed the existence of GPS in the linear array of 
threshold piecewise linear time-delay systems (Eq. (\ref{eq_3pl})) by using this concept of localized sets
as in the previous case.
Now, we will demonstrate the existence of GPS via sequential phase synchronization
in some randomly selected response systems ($i$ = 1, 10, 30, 50). 
The set obtained by sampling the time series of one of the systems whenever a maximum
occurs in the other system is plotted along with the attractors of the same systems.
The set, indicated as filled circles, obtained by observing the drive system 
($i = 1$) whenever the maxima occurs in the response system ($i = 10$) is shown in Fig.~\ref{fig15}(a) 
and that obtained by observing the response systems $i = 10,30,50$ whenever 
the maxima occurs in the drive system are shown in Figs.~\ref{fig15}(b-d) 
for the value of coupling strength $\varepsilon$ = 0.0. As the sets are spread over
the attractors, all the systems evolve independently and there is no CPS in 
the absence of coupling between them. Further when we increase the coupling
strength to $\varepsilon$ = 0.7, the oscillator ($i$ = 10) is partially synchronized with the drive
as the sets are almost localized but the sets in the oscillators $i$ = 30 and 50
are spread over the attractor which means that they are not yet phase synchronized 
with the drive system. This is shown in Figs.~\ref{fig15}(e-h). Again increasing the 
coupling strength to $\varepsilon$ = 0.85, the sets are further bounded to a small region over the 
attractors which shows that the oscillator $i$ = 10 is synchronized with the drive, but 
the oscillator $i$ = 30 is partially synchronized where the sets are almost localized 
and $i$ = 50 is not yet phase synchronized with the drive 
as represented by the spread of the sets over the attractor in Figs.~\ref{fig15}(i-l). 
Further, the Figs.~\ref{fig15}(m-p) and Figs.~\ref{fig15}(q-t) indicate the situation for 
$\varepsilon$ = 1.0 and $\varepsilon$ = 1.2, respectively, where all the oscillators 
are now phase synchronized with the drive as 
the sets are localized over the attractor confirming the existence of GPS in an array via 
sequential phase synchronization as the coupling strength is increased. It is to be noted that
the sets are not yet completely localized in Figs.~\ref{fig15}p for $\varepsilon = 1.0$, whereas
it is localized for $\varepsilon = 1.2$ (Figs.~\ref{fig15}l).
\begin{figure*}
\centering
\includegraphics[width=1.0\columnwidth]{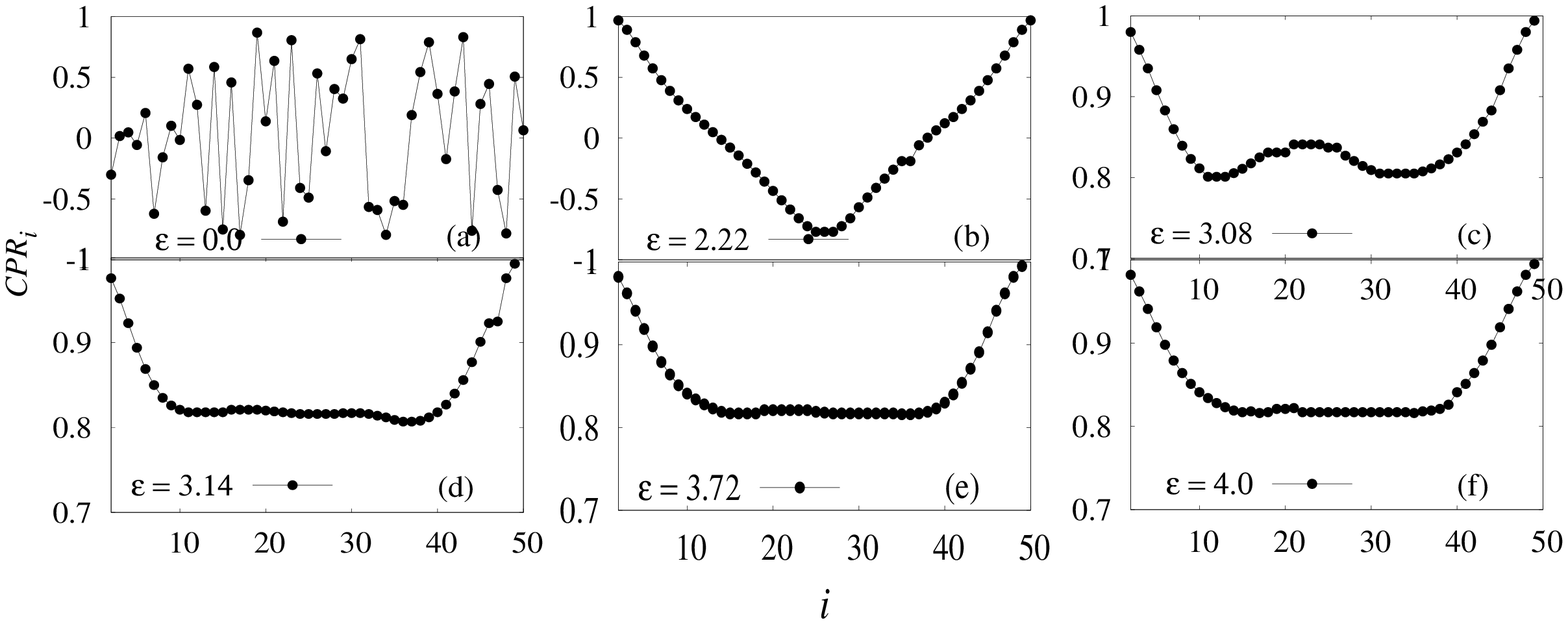}
\caption{\label{fig16} The index CPR as a function of the system index $i$ for 
various values of coupling strength $\varepsilon$ in ring topology for (\ref{eq_5}). (a) For $\varepsilon=0.0$, 
(b) for $\varepsilon=2.22$, (c) for $\varepsilon=3.08$,
(d) for $\varepsilon=3.14$, (e) for $\varepsilon=3.72$ and (f) for $\varepsilon=4.0$.}
\end{figure*}
\begin{figure*}
\centering
\includegraphics[width=1.0\columnwidth]{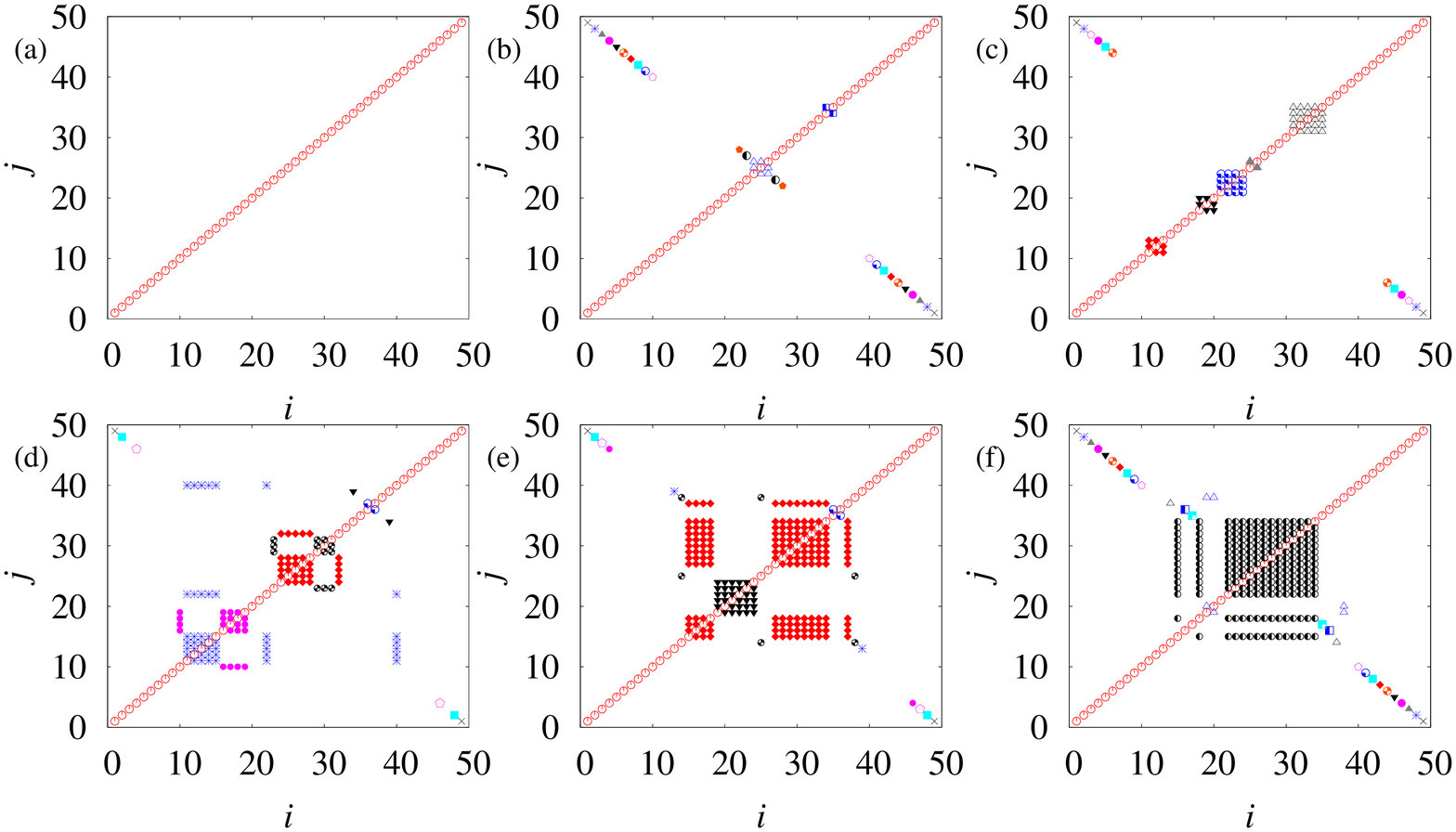}
\caption{\label{fig17} Snap shots of the node vs node plots for various values of coupling strengths
showing the occurrence of phase clusters. (a) For $\varepsilon=0.0$, 
(b) for $\varepsilon=2.22$, (c) for $\varepsilon=3.08$,
(d) for $\varepsilon=3.14$, (e) for $\varepsilon=3.72$ and (f) for $\varepsilon=4.0$.}
\end{figure*}

\section{\label{sec:level4}Partial Phase Synchronization (PPS) in a Linear Array of Piecewise Linear Time-delay Systems with Ring Topology}
So far we considered the dynamics of coupled arrays of time-delay systems with open end boundary conditions.
Now, we wish to demonstrate the existence of a partial phase synchronization (PPS)
in an array of time-delay systems with a ring topology. In this coupling configuration, all the systems are not
fully synchronized, instead they split into several subgroups to form 
different phase synchronized clusters. The phenomenon may also be called cluster synchronization. Here
the systems within each cluster maintain perfect phase synchronization. This kind of PPS
mostly occurs in neural networks \cite{danzl08}, chemical oscillations 
\cite{kiss02} and El-Ni$\tilde{n}$o systems \cite{karl11},
and has been studied in coupled chaotic systems \cite{bjorn06} as well. Also the transition from nonsynchronization
to phase synchronization via PPS has been studied in two-dimensional 
coupled map lattices \cite{zhuang02}. 

The dynamical equation of a linear array of coupled time-delay systems with closed end boundary conditions is given as
\begin{equation}
\dot{x}_i(t)=-\beta x_i(t)+ \alpha_{i} f(x_{i}(t-\tau))+ \varepsilon(x_{i-1}(t)-x_{i}(t)),
\label{eq_5}
\end{equation}
where $i=1,2,\cdots,N$, with the following periodic boundary conditions: 
$x_{0}=x_{N}$ and $x_{N+1}=x_{N}$. The function $f(x)$, and the system parameters have been chosen as in 
Sec.~\ref{sec:level2a}. In this type of coupling configuration the signal of the $N$th system is fed into
the first system and so there is no specific drive system and that each system sends its signal to the 
nearby system in a unidirectional way. In the absence of the coupling, there is no synchronization among the 
systems. If we increase the coupling, the systems with the 
same phase/frequency form a small group of clusters. 
If we increase the coupling above a threshold 
value ($\varepsilon_{thr}$), PPS occurs with a relatively large group of 
phase clusters. For even larger values, one finds that the PPS state collapses.
\begin{figure*}
\centering
\includegraphics[width=0.5\columnwidth]{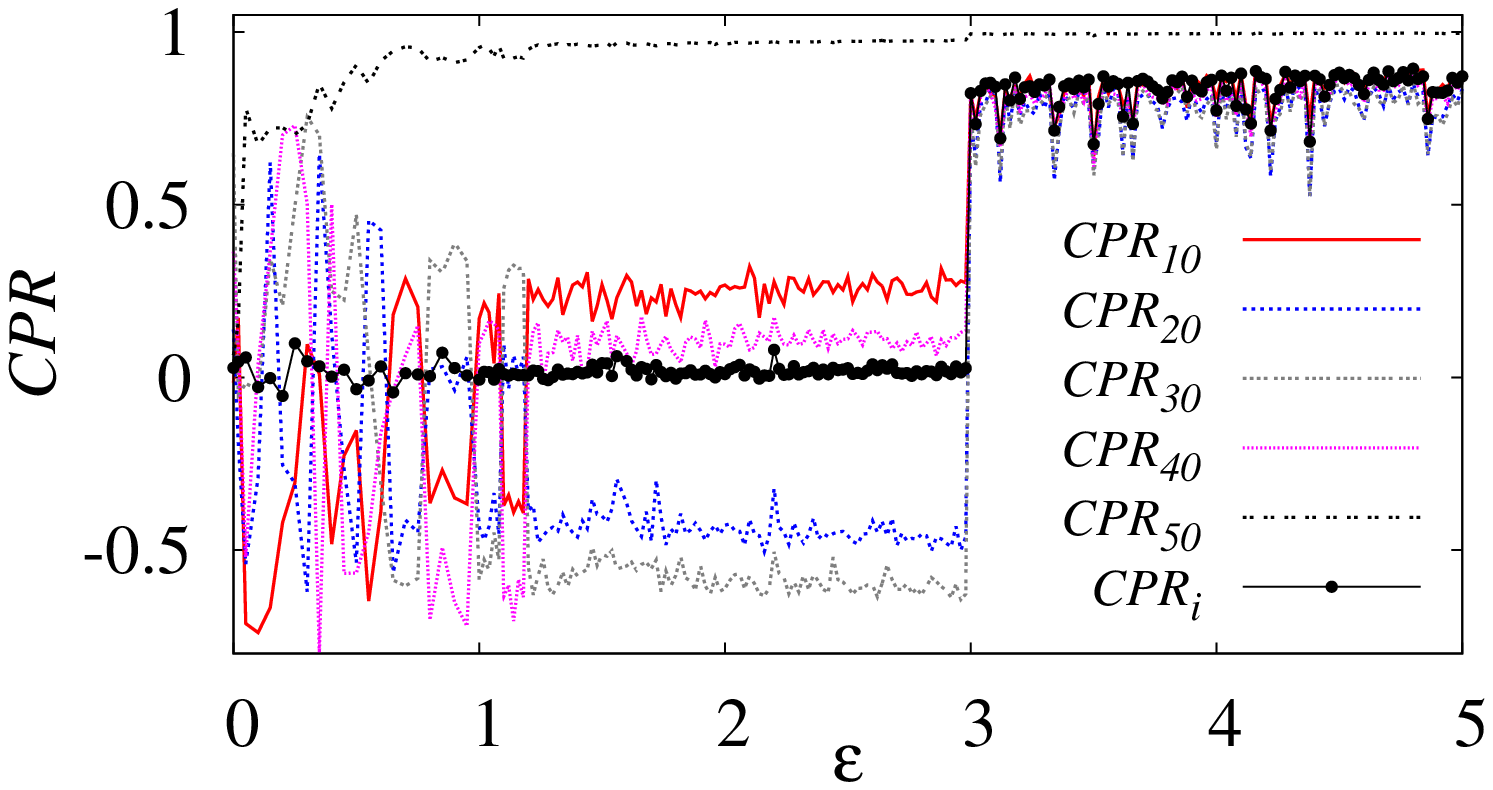}
\caption{\label{fig18} The index CPR as a function of the coupling strength $\varepsilon$. 
Different lines correspond to the CPR of different ($i=10,20,30,40$ and $50$)  
systems. The filled circles correspond to the mean value of the CPR of all the 
($N-1$) piecewise linear systems in the array.}
\end{figure*}

The nonlinear transformation used in Sec.~\ref{sec:level2} is no longer valid to 
transform the non-phase-coherent attractor into a phase-coherent attractor due to the 
new boundary condition. However, we have already found that the recurrence-based indices
serve as excellent quantifiers in identifying phase synchronization in
non-phase-coherent attractors both qualitatively and quantitatively. Further, we also
characterized the occurrence of the cluster formation using the concept of localized
sets. We will use these quantifications in the following.
\begin{figure*}
\centering
\includegraphics[width=1.0\columnwidth]{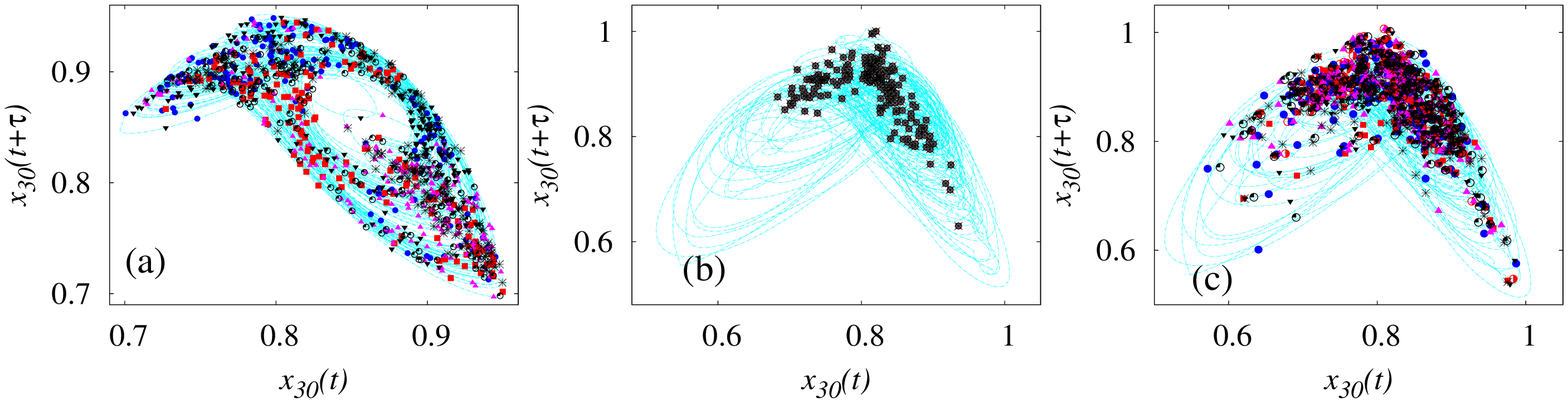}
\caption{\label{fig19} The occurrence of phase clusters is also 
explained using the concept of localized sets. (a) For $\varepsilon=0.0 $ figure shows
the attractor of the system $i=30$ and the sets of the systems $i=17, 18, 19, 22$ and $31$,
(randomly chosen) (b) for $\varepsilon=3.72$ with the sets of the systems inside the cluster 
$i=17, 28, 31, 34$ and $37$ (which lie within the cluster). (c) for $\varepsilon=3.72$ with the sets of the
systems outside the cluster $i=18, 19, 22$ and $27$.}
\end{figure*}

We have calculated the CPR of each of the $(N-1)$ systems
in the array by taking $i=1$ as the reference system (but we also confirmed that the
dynamics does not change even if we take any other system in the array as a reference system).
In Fig.~\ref{fig16} we have plotted the CPR of every system in the array as a function of 
the system index ($i$) for different values of coupling strength. In the absence of the 
coupling ($\varepsilon=0.0$), the systems in (\ref{eq_5}) evolve independently as indicated
by the random distribution of CPR in Fig.~\ref{fig16}(a). Increasing the coupling strength
to $\varepsilon=2.22$ results 
in synchronous evolution of a few of the oscillators, leading to groups of small clusters 
(which are having the same CPR values) as seen in Fig.~\ref{fig16}(b). However there is no 
single global synchronized state in the present network in contrast to the sequentially synchronized single
cluster as discussed in Sec.~\ref{sec:level2}. Above a threshold value 
($\varepsilon_{thr} > 3.0$), we can observe the occurrence of a PPS in the array
with large groups of phase synchronized cluster. Figs.~\ref{fig16}(c-f) indicate the occurrence of the 
PPS with separate groups of phase synchronized clusters for $\varepsilon=3.08$, $3.14$, $3.72$ and $4.0$, respectively.
If we increase the coupling strength for even larger values, PPS continues to exist until $\varepsilon$ increases to very 
large values ($\varepsilon>50$) where it collapses. For very large values of the coupling strength the clusters
lose their stability resulting in a desynchronized array. 
\begin{figure*}
\centering
\includegraphics[width=1.0\columnwidth]{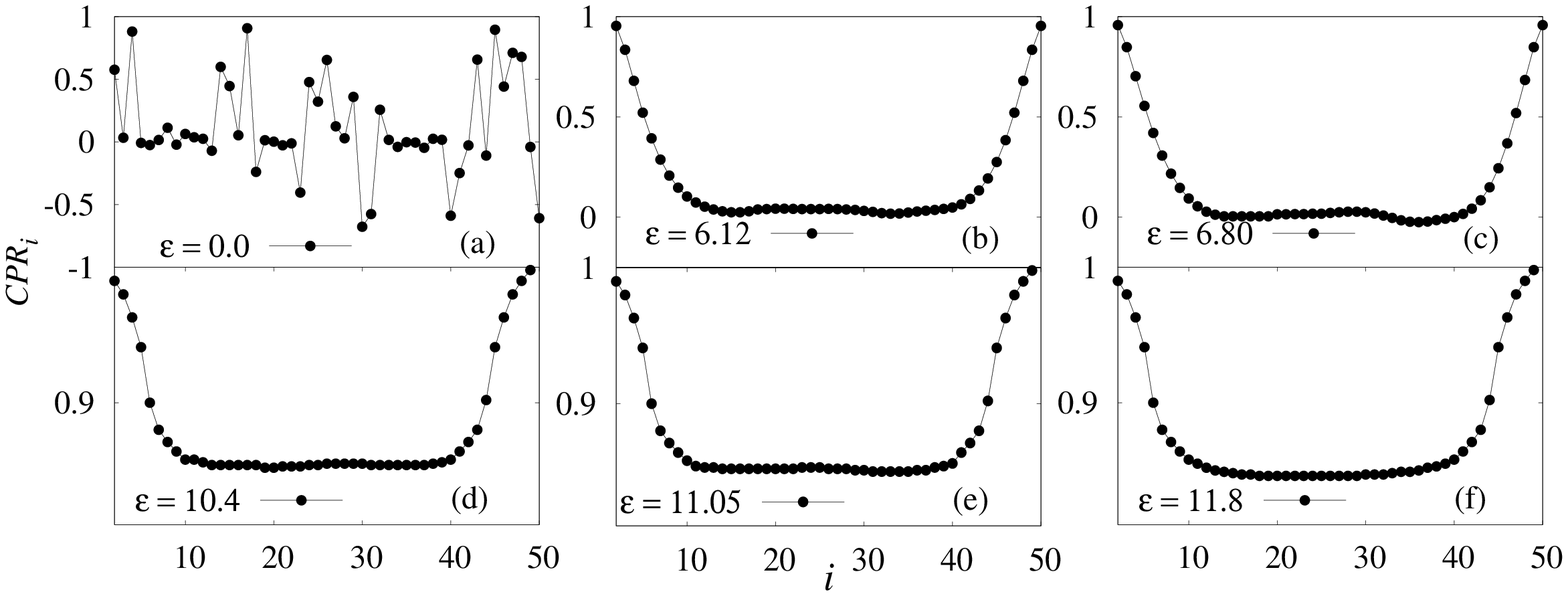}
\caption{\label{fig20} The index CPR as a function of the system index $i$ for 
various values of coupling strength $\varepsilon$ in the array (\ref{eq_5}) with ring topology of the threshold
piecewise linear system. (a) For $\varepsilon=0.0$, 
(b) for $\varepsilon=6.12$, (c) for $\varepsilon=6.80$, (d) for $\varepsilon=10.4$,
(e) for $\varepsilon=11.05$, and (f) for $\varepsilon=11.8$.}
\end{figure*}
\begin{figure*}
\centering
\includegraphics[width=1.0\columnwidth]{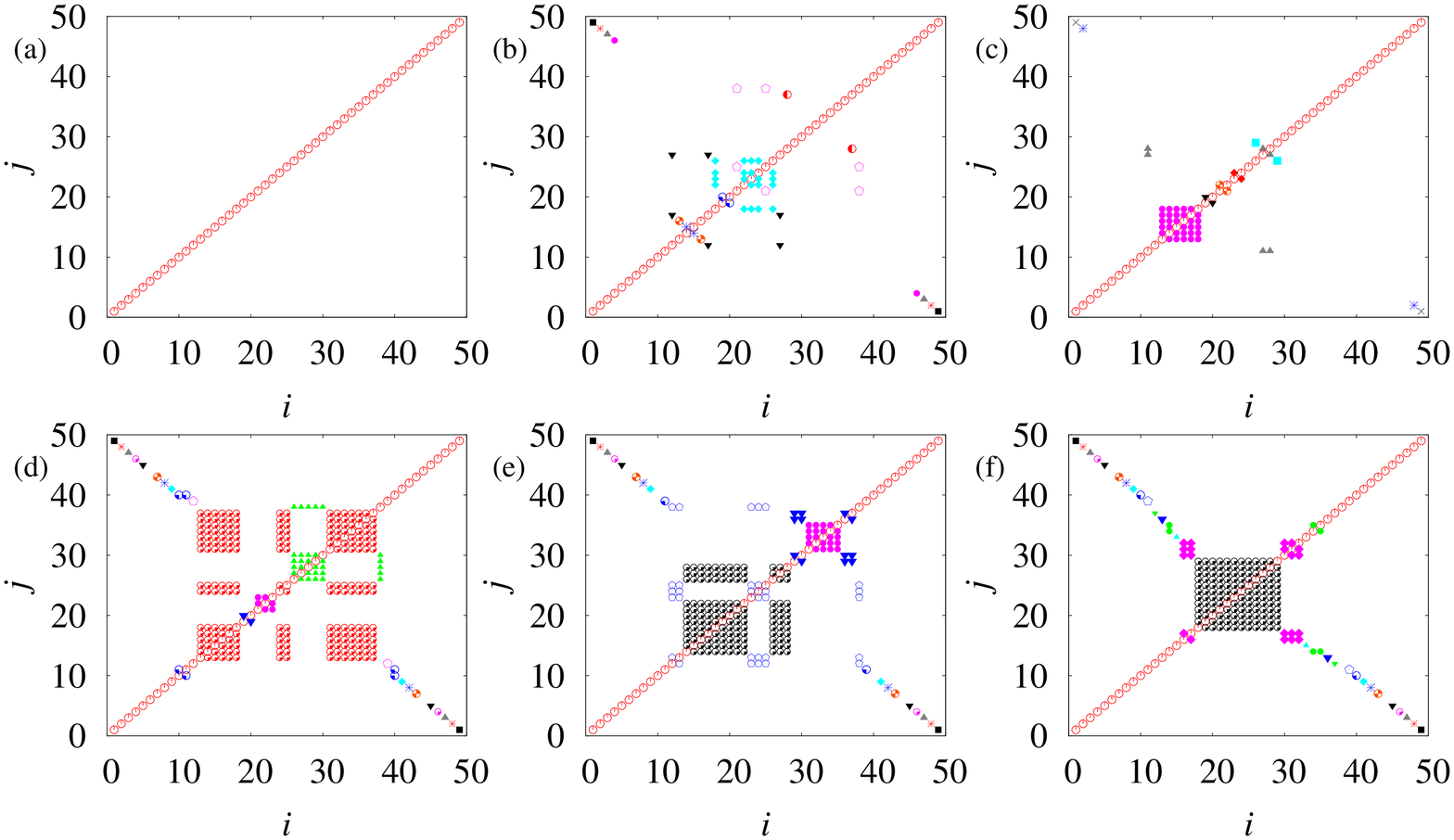}
\caption{\label{fig21} Snap shots of the node vs node plots for various values of coupling strengths
shows the occurrence of the phase clusters in threshold piecewise linear systems. (a) For $\varepsilon=0.0$, 
(b) for $\varepsilon=6.12$, (c) for $\varepsilon=6.80$, (d) for $\varepsilon=10.4$,
(e) for $\varepsilon=11.05$, and (f) for $\varepsilon=11.8$.}
\end{figure*}

The formation of the phase synchronized clusters can also be clearly illustrated and 
visualized by using the snap shots of the oscillators in the index vs index plot as 
shown in Fig.~\ref{fig17}. In this figure the systems with identical phase
(having the difference in their CPR value less than $0.001$) are assigned with identical shapes (colors). The diagonal line in 
Fig.~\ref{fig17}(a) for $\varepsilon=0.0$ correspond to the oscillator index $i=j$
and the oscillators evolve independently. Fig.~\ref{fig17}(b) displays
the occurrence of small groups of phase synchronized clusters in distant nodes ($i=2$ and $50$)
for $\varepsilon=2.22$. If the coupling is increased further 
the nodes with the same frequency form groups of clusters with each 
cluster having different phases resulting in PPS.
In contrast, in Fig.~\ref{fig5} due to the open end boundary
conditions, the nearest subsystems to the drive start to synchronize with it along with the 
formation of small groups of clusters. Further increase in the coupling strength results in the 
decomposition of the small clusters which then synchronize with the main cluster. Consequently GPS results 
in the array with open end boundary conditions. Figs.~\ref{fig17}(c-f)
show the occurrence of PPS in the network with the formation of large phase 
clusters for $\varepsilon=3.08, 3.14$, $3.72$ and $4.0$ respectively.

The mechanism for the formation of partial phase synchronization can be explained as follows: In the 
absence of the coupling the individual systems oscillate independently
with different phases due to the mismatches in the nonlinear parameters $\alpha_{i}$. Upon increasing the 
coupling strength, due to the ring configuration, every individual oscillator starts to 
drive the nearest oscillator and the oscillators with small differences in their phases/frequencies in the array 
synchronize themselves to form small groups of clusters leaving the other oscillators 
with large phase differences to evolve independently. Further increase in the coupling strength 
results in an increase in the sizes of the clusters due to the locking of the phases of the nearby
oscillators to the cluster for appropriate $\varepsilon$ and finally ending up with a large cluster near 
the center of the ring for large $\varepsilon$. However there exists other small clusters in the 
ring contributing to PPS and hence 
there is no GPS in such an array. As we have seen earlier in Sec.~\ref{sec:level2}, in GPS, 
as long as the coupling increases, the nearest subsystems of the drive 
start to synchronize with it and the remaining asynchronous systems
with the closest frequencies form small groups of phase clusters. Further increase in the 
coupling results in the formation of a single large cluster with the drive due to the decomposition
of the small clusters with different frequencies and thereby resulting in GPS. But in the ring configuration, if the 
coupling increases further (above the threshold value) the distant nodes with
same phase/frequency form groups of clusters with each cluster having its own phase resulting 
in PPS.

The synchronization transition to PPS is quantified using the value of CPR as a function
of the coupling strength. In Fig.~\ref{fig18} the different lines correspond to the
CPR of the randomly selected systems in the network ($i=10,20,30,40$ and $50$).
The filled circles correspond to the mean value of the CPR of all the ($N-1$) piecewise linear
systems. It is evident from this figure that only very few systems get phase synchronized  
to form small clusters and globally there is no phase synchronization when the coupling is below the threshold
value ($\varepsilon_{thr}<3.0$) which is indicated 
by the mean value of CPR near to zero. Once the coupling strength crosses the 
threshold value ($\varepsilon\approx3.0$), there is a sudden jump to CPR $\approx 0.8$
which is an indication of the occurrence of PPS. It is evident from this figure that the 
system $i=50$ is already synchronized with the drive reference system $x_{1}(t)$
as it is one of the nearest neighbors, which is also clearly seen in Figs.~\ref{fig16} and \ref{fig17}. 
One may note the interesting fact that this transition
is of first order (sharp) type, while in the case of open end boundary condition (see Sec.~\ref{sec:level2})
it is smooth and of second order type. The reason for this kind of transition is as follows:
In open end boundary conditions, the subsystems are sequentially synchronized with the drive, so the
GPS increases gradually in the array as a function of the coupling strength. 
Hence the transition is continuous (second order). But
in a ring topology all the systems attain PPS suddenly at a particular value of the coupling strength
due to the mutual sharing of the signals which leads to the discontinuous (first order) transition.

\begin{figure*}
\centering
\includegraphics[width=0.5\columnwidth]{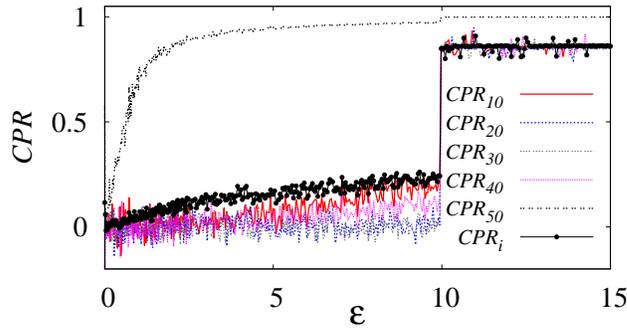}
\caption{\label{fig22} The index CPR as a function of the coupling strength $\varepsilon$
of the threshold piecewise linear time-delay system. 
Different lines correspond to the CPR of different ($i=10,20,30,40$ and $50$)  
systems. The filled circles correspond to the mean value of the CPR of all the 
($N-1$) piecewise linear systems 
in the array.}
\end{figure*}

Next we calculate
the localized sets by defining the event among the systems within the cluster and observing
the other systems that are in the same cluster. In this case the obtained sets of the 
systems within the cluster are localized on the attractor, while the sets of the systems 
outside the cluster gets spread over the entire attractor. For illustration, in Fig.~\ref{fig19}(a)
the attractor of the system $i=30$ is plotted along with the sets of the randomly selected
systems $i=17,18,19,22$ and $31$ (which are represented by different shapes and colors) where the sets are
spread over the attractor which corresponds to the unsynchronized state for $\varepsilon=0.0$.
Fig.~\ref{fig19}(b) corresponds to $\varepsilon=3.72$ for the sets of the 
systems $17,28,31,$ and $37$ which are all within the cluster. The figure shows that the sets of the corresponding systems 
are localized on the attractor in the same place implying that these systems form 
a clustered state. On the other hand, for the same value of the coupling, we have plotted the sets 
of the systems outside the cluster ($i=18,19,22$ and $27$).  
This shows that the sets are spread over the entire attractor indicating that the corresponding systems
are not in phase synchronization (Fig.~\ref{fig19}(c)).

\section{\label{sec:level5}Partial Phase Synchronization in an Array of Piecewise Linear Systems with Threshold Nonlinearity}
In this section, we demonstrate the existence of the PPS in the second piecewise linear
time-delay system with a threshold nonlinear function (\ref{eq_3pl}). The form of the piecewise function 
$f(x)$ in Eq.~(\ref{eq_5}) and the system parameters are set to be the same as in Sec.~\ref{sec:level4}. 

In Fig.~\ref{fig20}, the CPR is shown as a function of the system
index ($i$) for different values of the coupling strength. For $\varepsilon=0.0$, in Fig.~\ref{fig20}(a), there is no
correlation in the values of CPR of the systems attributing to the desynchronization
state. Figures~\ref{fig20}(b) and \ref{fig20}(c) correspond to the coupling strength
$\varepsilon=6.12,$ and $6.80$, respectively, which display the occurrence of small 
phase synchronized clusters. If we increase the coupling strength $\varepsilon>10.0$ 
one can observe the occurrence of PPS in the array with large 
groups of phase synchronized clusters. Figures~\ref{fig20}(d-f) indicate the occurrence of the
PPS with separate groups of phase synchronized clusters for $\varepsilon=10.4, 11.05$ and $11.8$,
respectively.

The dynamical organization of the cluster formation is clearly visualized in Fig.~\ref{fig21}, node vs
node plot, for various values of the coupling strength. The different shapes (colors) correspond to the nodes 
which are in the different phase synchronized clusters. In Fig.~\ref{fig21}(a) the diagonal line shows the oscillator index
$i=j$ evolving independently for $\varepsilon=0.0$. Figures~\ref{fig21}(b) and \ref{fig21}(c)
display the occurrence of several small phase synchronized clusters for $\varepsilon=6.12$ 
and $6.80$, respectively. If the coupling is increased further, the nodes with the same frequency form groups
of clusters with each cluster having different phases resulting in PPS as shown in Fig.~\ref{fig21}(d-f)
for the values of $\varepsilon=10.4, 11.05$ and $11.8$, respectively.
\begin{figure*}
\centering
\includegraphics[width=1.0\columnwidth]{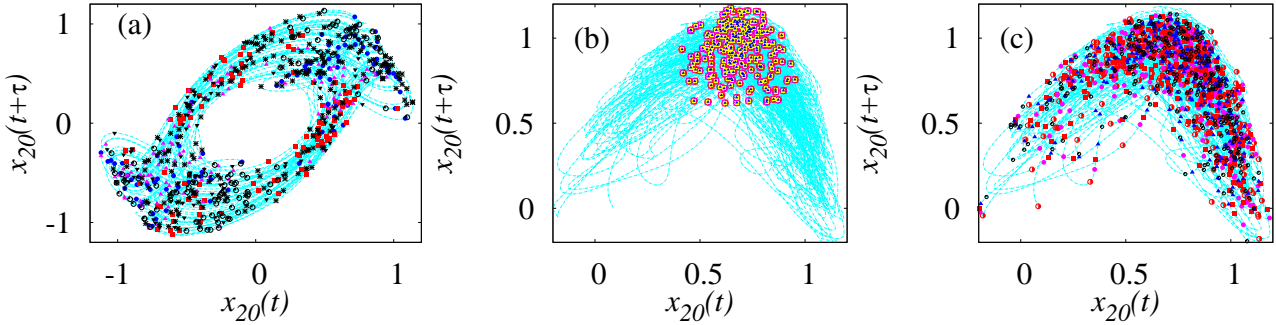}
\caption{\label{fig23} The occurrence of the phase clusters in threshold piecewise linear systems
in the ring topology
is explained using the concept of localized sets. (a) For $\varepsilon=0.0 $ shows
the attractor of the system $i=20$ and the sets of the systems $i=10,15,19,18, 24$ and $40$,
(b) for $\varepsilon=11.05$ with the sets of the systems inside the cluster 
$i=15,16,18$ and $26$. (c) for $\varepsilon=11.05$ with the sets of the
systems outside the cluster $i=10,13,24$ and $40$.}
\end{figure*}

The transition from nonsynchronization to partial phase synchronization is quantified
by plotting the CPR as a function of the coupling strength. In Fig.~\ref{fig22} the different lines 
correspond to CPR of the randomly selected systems ($i=10,20,30,40$ and $50$) and the filled circles 
represent the mean value of the CPR of ($N-1$) systems. It is evident from this figure
that there is no synchronization (CPR$<0.2$) below $\varepsilon<10.0$.
There is a sudden jump (a first order transition) in the value of CPR (to $\approx 0.85$)  for $\varepsilon \approx 10.0$
indicating the onset of PPS in the array along with the formation of a large group of 
phase synchronized clusters.

The occurrence of these phase synchronized clusters in the array of threshold piecewise linear
time-delay systems is also demonstrated using the concept of localized sets. In Fig.~\ref{fig23},
we have plotted the attractor of the system $i=20$ along with the sets of some randomly selected
systems in the array. Each shapes (colors) correspond to the sets of different
subsystems. Figure.~\ref{fig23}(a) is plotted for $\varepsilon=0.0$ and the sets of the systems
$i=10,15,18,24$ and $40$ spread over the entire attractor indicating the
asynchronized state. For $\varepsilon=11.05$, the sets of the systems ($i=15,16,18$ and $26$) inside the large cluster 
 are localized on the attractor as shown in 
Fig.~\ref{fig23}(b) illustrating that the respective systems are in a phase synchronized state.
On the other hand, for the same value of the coupling strength, we have plotted the sets of the systems outside 
the cluster ($i=10,13,24$ and $40$), which are spread over the attractor $i=20$ 
indicating that these subsystems are not in a phase synchronized state (see Fig.~\ref{fig23}(c)).

\section{\label{sec:level6}Summary and Conclusion}

We have demonstrated the existence of global and partial phase synchronizations in an array
of unidirectionally coupled nonidentical piecewise linear time-delay systems with two different boundary conditions.
Coupled with our earlier work on Mackey-Glass system \cite{suresh10}, our studies clearly establish
the generic nature of the underlying phenomena. 
In particular, in a linear array with open end boundary conditions, we have demonstrated the emergence 
of GPS via sequential synchronization.
We have shown that in addition to the main phase synchronized cluster centered at the drive,
the remaining asynchronous systems from the main cluster organize themselves to form 
different clusters for low values of the coupling strength. Further increase in the coupling strength
leads to the formation of a single large cluster resulting in GPS by a decomposition of the
other clusters. The synchronization transition is of second order type.
We have confirmed the existence of GPS
via sequential phase synchronization by estimating the phase difference, the average 
frequency and the average phase as a function of the oscillator index and the
coupling strength from the transformed 
attractors. Furthermore, we have also confirmed the existence of GPS
from the original non-phase-coherent attractors of the coupled piecewise linear time-delay systems
using appropriate recurrence quantification measures and the concept of localized sets 
without explicitly estimating the phase. We have also confirmed the existence of 
GPS via sequential phase synchronization in a second piecewise linear 
system with threshold nonlinear function.

The existence of a partial phase synchronization (PPS) is demonstrated in an array with 
closed end boundary conditions (ring topology) and the synchronization transition is of first order type.
 PPS is corroborated using recurrence 
analysis and the concept of localized sets using the
original non-phase-coherent hyperchaotic attractor of the piecewise linear time-delay
systems. Further, the mechanism for the formation of both GPS and PPS are elaborated. 
We have also confirmed the existence of PPS in an array
of Mackey-Glass time-delay systems with closed end boundary conditions; however 
the results are not given here as they are similar in nature to the above systems.

This study can also be extended to two-dimensional lattices of coupled
time-delay systems with and without delay coupling and in networks of 
time-delay systems with indirect global dynamic environment coupling, 
as well as networks with complex topology.

\section*{Acknowledgments}
The work of R.S. and M.L. has been supported by the Department of Science 
and Technology (DST), Government of India sponsored IRHPA research project.
M.L. has also been supported by a DST Ramanna project and a DAE Raja Ramanna Fellowship.
D.V.S. and J.K. acknowledge the support from 
EU under project No. 240763 PHOCUS(FP7-ICT-2009-C).

\end{document}